\documentclass[useAMS,usenatbib]{mn2e}
\usepackage{epsfig}
\usepackage{graphicx}
\usepackage{amsmath}

\title[Interpreting broad emission-line variations II]{
Interpreting broad emission-line variations II: \newline 
Tensions between luminosity, characteristic size \newline and responsivity}
\author[Mike,
  Kirk]{M. R. Goad$^{1}$\thanks{E-mail: mg159@le.ac.uk} and
  K. T. Korista$^{2}$\\
$^{1}$Department of Physics and Astronomy, 
College of Science and Engineering, University of Leicester,  University Road, Leicester, LE1 7RH\\
$^{2}$Department of Physics, Western Michigan University,  Kalamazoo, Michigan 49008-5252, USA\\}

\begin{document}
\date{Received xxx; in original form
  June 2015}
\pagerange{\pageref{firstpage}--\pageref{lastpage}} \pubyear{2015}
\maketitle
\label{firstpage}

\begin{abstract}

 We investigate the variability behaviour of the broad H$\beta$
  emission-line to driving continuum variations in the best-studied
  AGN NGC~5548. For a particular choice of BLR geometry, H$\beta$
  surface emissivity based on photoionization models, and using a
  scaled version of the 13~yr optical continuum light curve as a proxy
  for the driving ionizing continuum, we explore several key factors
  that determine the broad emission line luminosity $L$,
  characteristic size $R_{\rm RW}$, and variability amplitude (i.e.,
  responsivity) $\eta$, as well as the interplay between them.

For fixed boundary models which extend as far as the hot-dust the
predicted delays for H$\beta$ are on average too long. However, the
predicted variability amplitude of H$\beta$ provides a remarkably good
match to observations except during low continuum states. We suggest
that the continuum flux variations which drive the redistribution in
H$\beta$ surface emissivity $F(r)$ do not on their own lead to large
enough changes in $R_{\rm RW}$ or $\eta_{\rm eff}$.  We thus
investigate dust-bounded BLRs for which the location of the effective
outer boundary is modulated by the continuum level and the
dust-sublimation and dust-condensation timescales. We find that in
order to match the observed variability amplitude of broad H$\beta$ in
NGC~5548 a rather static outer boundary is preferred.

Intriguingly, we show that the most effective way of reducing the
H$\beta$ delay, while preserving its responsivity and equivalent
width, is to invoke a smaller value in the incident ionizing photon
flux $\Phi_{\rm H}$ for a given ionizing source--cloud radial distance
$r$, than is normally inferred from the observed UV continuum flux and
typical models of the continuum SED.

\end{abstract}

\begin{keywords}
galaxies: active -- galaxies: Seyfert -- quasars: emission lines -- methods: numerical
\end{keywords}

\section{Introduction}

Time variable continuum and broad emission-line studies (reverberation
mapping) have demonstrated beyond doubt that the broad emission line
region (hereafter BLR) in Active Galactic Nuclei (AGN) is both
geometrically thick and highly stratified, with strong gradients in
density and/or ionization state (e.g. Krolik et al. 1991; Clavel et
al. 1991; Peterson et al. 2002, and references therein). Observations
of multiple UV and optical broad emission-lines in individual sources
and which span a range in ionization potential indicate that the high
ionization lines (HILs, e.g. N~{\sc v}, C~{\sc iv}, He~{\sc ii})
respond on the shortest timescales, while the low ionization lines,
e.g. Mg~{\sc ii}, Fe~{\sc ii} and optical/IR recombination lines
(H$\alpha$, H$\beta$, H$\gamma$, Pa~$\alpha$) respond on longer
timescales (e.g., Peterson et al. 2002; Barth et~al. 2013). If the
continuum--emission-line delays relate simply to the separation
between the continuum and emission-line forming regions, then the
observed differences between the response timescales of individual
lines suggest that lines of differing ionization potential
preferentially form at different radial distances (implying a
spatially extended BLR), with the HILs forming closer to the continuum
source than the LILs.

More recently, dust reverberation studies place the dust emission just
beyond the furthest reaches of the BLR, with delays somewhat larger
than the largest delays measured for the LILs (Suganuma et al. 2004,
2006; Koshida et al. 2009, 2014; Schn\"{u}lle et al. 2013, 2015; Kishimoto
et~al. 2013). At smaller radii, the intense radiation field may
contribute to cloud destruction (Mathews 1982), over-ionize the gas,
or the surviving clouds may become continuum sources in this high
pressure environment (Ferland and Rees 1988).  Thus a picture of
the BLR is emerging in which the BLR gas is bounded on its inner edge
by continuum sources (i.e. the inner accretion disc), while at large radii
the extinction of ionizing photons and the destruction of
optically-thick emission lines by grains, causes the efficiency at
which a particular emission-line forms to drop significantly
(e.g. Netzer and Laor 1993).

For NGC~5548, in order to match the observed luminosities of the
strong UV lines, photoionization model computations indicate that its
BLR spans $\sim$2 decades in radial extent from
$\sim$1--100~light-days (Korista and Goad 2000; Kaspi and Netzer
1999).

While photoionization models alone can broadly constrain the radial
extent of the BLR, information concerning the spatial distribution and
kinematics of the BLR gas within the confines of these boundaries
requires the additional information supplied by reverberation mapping
of multiple broad emission-lines in individual sources.  The most
recent emission-line velocity--delay maps indicate evidence for
inflow, outflow and circularised motion, often with evidence for more
than one type of motion in a single source (Skielboe et al. 2015;
Grier et al. 2013;  Pancoast et al. 2014b; Barth et al. 2011; Bentz et
al. 2010; Denney et al. 2009, 2010).  The prevailing view is that gas
motion is generally virialised, a fact which has been usefully
exploited in conjunction with BLR `size' estimates to determine black
hole masses for approaching nearly 60 nearby low luminosity AGN (see
Bentz and Katz 2015 for the most recent compilation of reverberation
mapped sources; Peterson et al. 2004, and references therein; Bentz
et~al. 2008, 2009a,b; Pancoast et al. 2012, 2014a, 2014b).  However,
the spatial distribution of the BLR gas is far harder to fathom.  The
1-d response function $\Psi^{\prime}(\tau)$, the function which maps
the continuum variations on to the emission-line variations, alone is
degenerate, with disparate geometries yielding broadly similar 1-d
response functions (Welsh and Horne 1991; P\'{e}rez, Robinson and de
la Fuente 1992), and the recovered 2-d response functions
$\Psi^{\prime}(v,\tau)$ (or velocity--delay maps) are only now
reaching sufficient fidelity to provide useful constraints on the gas
distribution and kinematics (Grier et al. 2013; Pancoast et al. 2012,
2014a, 2014b; Bentz et al. 2010).

In this work, we explore the role of the BLR outer boundary in
establishing an emission line's luminosity, delay (lag), and
variability amplitude (responsivity) in the presence of ionizing
continuum variations, and identify the strong connection between these
three quantities. The most up to date measurement of the
size-luminosity relation for broad H$\beta$ in nearby AGN suggests a
relation of the form $R_{\rm BLR}({\rm H}\beta) \propto L_{\rm
  opt}^{\alpha}$, with $\alpha \approx 0.5$ (Bentz et al. 2009; Bentz
et al. 2013; Kilerci Eser et al. 2015). This scaling is a naive prediction of
the simplest photoionization model calculations. The broad similarity
between AGN spectra spanning a wide range in continuum source
luminosity suggests that the physical conditions within the
line-emitting gas are broadly similar from one object to the next. In
that case, the ionization parameter $U$ which relates the number of
hydrogen ionizing photons $Q_{\rm H}$ to hydrogen gas number density
$n_{\rm H}$ and ionizing continuum source--cloud separation $r$
through the relation $U = Q_{\rm H}/4\pi r^{2} n_{\rm H}c$, then gives
$r \propto (Q_{\rm H} / Un_{\rm H})^{1/2}$.

\subsection{The BLR in NGC~5548 from correlated continuum--emission-line variability studies}
UV spectroscopic monitoring of the best studied source, the nearby
Seyfert 1 galaxy NGC 5548, reveals that the HILs undergo large
amplitude short timescale variations. Indeed, the 1-d response
functions for the HILs (N~{\sc v}, C~{\sc iv}, He~{\sc ii}) are
temporally unresolved on the shortest timescales, peaking at zero
delay, and declining rapidly toward longer delays, with a mean
response timescale of only a few days (e.g. Krolik et al. 1991; Clavel
et~al. 1991; Korista et al. 1995). By contrast, the amplitude of the
response for the LILs (Mg~{\sc ii}, Fe~{\sc ii}) is much weaker, and
their mean response timescales both larger and with large uncertainty
(Clavel et al.1991; Maoz et al. 1993; Vestergaard and Peterson 2005;
Kuehn et al. 2008; Barth et al. 2013).  Furthermore, the C~{\sc iii}]
  inter-combination line, a line which is collisionally de-excited at
  high gas densities, also displays a smaller amplitude response and
  longer delays than the HILs suggesting that densities generally
  decrease toward larger BLR radii in this source. The amplitude and
  response timescales for the broad optical recombination lines are
  intermediate between those of the HILs and LILs (e.g Peterson et
  al. 2002).  For example, for broad H$\beta$ the 1-d response
  function recovered from the 13~yr ground-based optical monitoring
  campaign on NGC~5548 is characterised by an absence of response on
  both short and long timescales, rising to a peak response on
  timescales of order 20~days, with a full-width at half-maximum
  (FWHM) of $\approx 10$ days (e.g. Horne, Welsh and Peterson 1991;
  Cackett and Horne 2006). See paper~{\sc i} for a more detailed
  description of the 13~yr ground-based monitoring data on NGC~5548.

The simplest explanation for the differences in the measured delays
for lines of differing ionization potential, is that the BLR in
NGC~5548 is spatially extended and highly stratified, though the
typical delays for the responding region appear to imply a far more
compact BLR than photoionization models might suggest, and which
remains a challenge in terms of balancing the energy budget for the
strongest UV and optical recombination lines (Netzer 1985,
Collin-. Souffrin 1986).  The absence of significant response for
broad H$\beta$ on the shortest timescales can be explained either by
an absence of gas along the line of sight to the observer, implying a
significant departure from spherical symmetry, for example a disc or
bowl-like geometry, or by line of sight gas which is very
optically-thick in the line, in which case the emission-line will
emerge predominantly from the illuminated face of the
cloud\footnote{Dynamical modelling of reverberation mapping data taken
  for the optical recombination lines in a small number of nearby
  Seyfert 1 galaxies, indicates that the observed line emission for broad
  H$\beta $ emerges preferentially from the illuminated face of the
  BLR clouds (e.g. Pancoast et~al. 2014a).}.

Ferland et al. (1992), O'Brien et al. (1994), and Korista and Goad
(2004) estimate that the fraction of Balmer line photons emerging from
the illuminated face of typical BLR clouds lies in the range
80--100\%.  However, only a bowl-shaped BLR geometry can
simultaneously account for the absence of significant response in the
line on both short and long timescales (Goad, Korista and Ruff
2012). For a bowl-shaped geometry, BLR gas is elevated above the disc
mid-plane and gives rise to smaller time-delays for a fixed radial
distance than would otherwise be expected for spherical or flattened
BLR geometries.  In the limit where gas lies along an iso-delay
surface, the measured delay is independent of cloud--ionizing
continuum source distance. For NGC~5548, a bowl-like model also has
the significant advantage in that it reconciles the measured distance
to the hot dust, as estimated from the delay between the optical and
IR continuum bands ($\tau \approx 50$ days, Suganuma et al. 2004) with
photoionization calculations, which predict a minimum distance of
$\approx 100$~light-days for grain survival.

\subsection{Factors influencing the emission-line responsivity}

Goad and Korista (2014, hereafter paper~{\sc i}) investigated the
effect of geometric dilution on the amplitude of the emission-line
response (the line responsivity $\eta_{\rm eff}$) and delay.
Formally, the line responsivity, is the power-law index which relates
the measured continuum and emission-line fluxes, $f_{\rm cont}$ and
$f_{\rm line}$, via

\begin{equation}
f_{\rm line} \propto f_{\rm cont}^{\eta_{\rm eff}}  \, .
\end{equation}

\noindent $\eta_{\rm eff}$ is a measurable quantity, and is normally
estimated after first applying a small correction for the
continuum--emission-line delays (e.g. Pogge and Peterson 1992), with
due allowance for contaminating galaxy and narrow emission-line
contributions to the continuum and broad emission-lines respectively.

The efficiency by which ionizing continuum photons are converted into
emission line photons, the emission line EW, is related to the line responsivity
according to

\begin{equation}
\frac { d \log EW } { d \log f_{\rm cont} } = \eta_{\rm eff} - 1  \,  .
\end{equation}

\noindent The connection between the line responsivity and the line EW
is thus made clear. For $\eta_{\rm eff}=1$, the $EW$(line)=$constant\/$
with respect to the change in the incident continuum, and thus the
reprocessing efficiency for a particular line is independent of (a
finite change in) continuum level. Values of $\eta_{\rm eff} < 1$,
indicate that the line reprocessing efficiency diminishes with
increased continuum flux (an intrinsic Baldwin effect for that line,
e.g. Gilbert and Peterson 2003; Korista and Goad 2004; Goad, Korista
and Knigge 2004; Goad and Korista 2014). While for $\eta_{\rm eff}>1$,
the line reprocessing efficiency increases with increasing continuum
flux.

In paper~{\sc i}, Goad and Korista (2014) demonstrated that an
alternative estimate of $\eta_{\rm eff}$ can be made using only the
ratio of the variances of the line and continuum light-curves, as was
first suggested by Krolik et al. (1991), once again after first
applying a suitable correction for the contaminating galaxy and
narrow-line components.  Factors controlling the measured
emission-line responsivity $\eta_{\rm eff}$ include : (i) the local
gas reprocessing efficiency (Korista and Goad 2004, hereafter referred
to as the local line responsivity $\eta(r)$), (ii) the amplitude and
characteristic timescale of the driving continuum light-curve, (iii)
the BLR geometry and observer line of sight orientation, and (iv) the
duration of the monitoring experiment.

Previous studies indicated that the emission-line response was only
weakly affected by geometric dilution (e.g. Gilbert and Peterson
2003). However, Goad and Korista (2014) showed that for geometries
approaching the size expected for NGC~5548, geometric dilution could
in fact be significant. In general the importance of geometric
dilution depends upon the characteristic timescale $T_{\rm char}$ of
the driving continuum light-curve in relation to the maximum delay of
a given emission-line $\tau_{\rm max}$(line). $\tau_{\rm max}$(line)
is here defined to be the maximum time-delay for a given emission line
for a particular geometry and observer line-of sight orientation.  For
$T_{\rm char} < \tau_{\rm max}$(line), the measured responsivity and
delay, directly correlate with $T_{\rm char}$.

To illustrate these effects, in paper~{\sc i} we modelled the radial
surface emissivity distribution for the line emitting gas as a simple
power-law in radius $F(r)\propto r^{\gamma}$, since under these
conditions the local radial line responsivity $\eta(r)$ is constant,
both spatially and temporally (i.e. $\eta(r) =-(\gamma/2)=constant$).
Therefore, in this case the only factors which can influence the measured line
responsivity are the campaign duration, amplitude and characteristic
timescale of the driving continuum light-curve for a particular choice
of BLR geometry and observer line-of-sight orientation. In all cases
broad emission-line light-curves were determined using a
locally-linear response approximation for a stationary BLR with fixed
boundaries.  Thus, the models explored in paper~{\sc i} were by design
constructed to be time-steady. As such, the amplitude of the
emission-line response and the responsivity-weighted size of the BLR
will remain constant in time.  Therefore, ``breathing'', the observed
positive correlation between BLR size and continuum state (Korista and
Goad 2004; Goad, Korista and Knigge 2004; Cackett and Horne 2006;
Bentz et al. 2007; Kilerci Eser et al. 2015), was not addressed.

Paper~{\sc i} also indicated that the measured emission-line
responsivity and delay for a fixed $T_{\rm char}$, show significant
variation for BLRs differing only in their radial surface emissivity
distribution and/or their spatial extent.  Thus, if either the local
surface emissivity distribution and/or the location of the BLR inner
and outer boundaries were to vary with continuum level, we would
expect to find significant changes in the measured responsivity and
delay.  For static BLR boundaries changes in the measured responsivity
and delay require radial surface emissivity distributions that depart
significantly from a simple power-law over the radial extent of the
BLR.  Breathing requires the emission-line responsivity to generally
increase toward lower incident ionizing continuum fluxes {\it or
  equivalently larger BLR radii\/}.  This ensures that the mean line
formation radius will drop in low continuum states and rise in high
continuum states as is observed. This requirement may be relaxed, if
instead the BLR boundaries are allowed to adjust with continuum level,
moving outwards/inwards as the continuum rises/falls (though not
necessarily in lock-step). In practice, it may be that both are
required to match the observed behaviour of the emission-lines in
response to continuum variations.

Investigating the redistribution of the emission-line energy within
the BLR, in the presence of incident continuum flux variations is the
subject of the following contribution.  In the context of model BLRs
with both static and varying boundaries (for a particular geometry and
observer orientation), we here explore the nature of the broad
emission-line response, its amplitude and delay, assuming a full
non-linear response in the lines, and using radial surface emissivity
distributions for the H$\beta$ emission line determined from
photoionization calculations. As we show in \S2, these indicate
significant departures from a simple powerlaw radial surface
emissivity distribution over the radial extent of the fiducial BLR
geometry. We compare model predictions to the 13-year light curve for
broad H$\beta$ in NGC~5548 (Peterson et al. 2002) to elucidate the
physical factors that determine a particular emission line's average
luminosity, its response delay (`lag') and amplitude (`responsivity')
-- as well as the relationship between these three quantities.

This paper is structured as follows: In \S2 we describe the fiducial
BLR geometry for NGC~5548, onto which we project a Locally
Optimally-emitting Clouds (hereafter LOC) model description for the
radial surface emissivity distribution $F(r)$, constructed from
photoionization calculations which use a continuum normalisation
appropriate for this source.  In \S3 we drive the fiducial model as
well as representative powerlaw models for the radial surface
emissivity distribution, to produce broad H($\beta$) emission-line
light curves which we compare to the broad H($\beta$) emission-line
lightcurve as observed over 13 years of ground-based monitoring of
NGC~5548.  In \S4 we examine the link between the emission line
luminosity, emission line responsivity and emission-line delay,
highlighting those physical affects which when present can act to
enhance the H$\beta$ luminosity and response amplitude at small BLR
radii.  In \S5 we use simple toy models of a dust-bounded BLR to
explore the nature of the BLR outer boundary, and its effect on the
emission-line response amplitude and delay.  In \S6 we demonstrate the
implications (in terms of the measured response amplitude and delay)
for a dust-bounded BLR model for NGC~5548.  We discuss and summarise
the results in \S7.

For the purposes of computing continuum and emission line luminosities
and determining physical radial scales within the BLR, we adopt the
redshift of NGC~5548, z=0.0172, and the following cosmological
parameters: $H_{\rm 0} = 67.3$~km~s$^{-1}$~Mpc$^{-1}$, $\Omega_{\rm m}
= 0.315$, $\Omega_{\Lambda}=0.685$, with a corresponding luminosity
distance to NGC~5548 of 77.6~Mpc (Ade et al. 2014).

\section{A fiducial BLR model}\label{fiducial}

The fiducial BLR geometry has been described in detail elsewhere
(e.g. Goad, Korista, and Ruff 2012; Goad and Korista 2014).  This
geometry was introduced as a means of connecting the outer accretion
disc with the reservoir of gas that likely fuels the continuum source
(i.e. the dusty torus). By introducing significant scale-height at
large BLR radii (see also Collin et al. 2006) this geometry can
reconcile the small dust-delays reported for NGC~5548 (Suganuma et
al. 2004, 2006; Koshida et al. 2014), with the larger predicted
distance at which robust graphite grains sublimate (Mor and Netzer
2012).  In summary, the broad line-emitting gas occupies the surface
of an approximately bowl-shaped region characterised in terms of a
radially dependent scale height $H$, according to

\begin{equation}
H=\beta( r_{x})^{\alpha} \;  ,
\end{equation}

\noindent where $r_{x}$ is the projected radial distance along the
plane perpendicular to an observers line of sight (i.e. $r_{x} = r
\sin \phi$, $r$ is the cloud source distance, $\phi$ is the angle
between the polar axis and the surface of the bowl), and $\alpha$,
$\beta$ control the rate at which $H$ increases with
$r,\phi$.\footnote{The bowl-shaped surface is a zeroth order
  approximation of a BLR geometry in which the scale height is
  significant and increases with increasing radial distance. We do not
  exclude the possibility that line emitting gas exists ``behind'' the
  bowl surface (which may itself be patchy) at large physical depths,
  and which sees an extinguished ionizing continuum. The subsequent
  escape of such lines  will depend upon the cloud distribution.}
We choose a velocity field of the form
\begin{equation}
v_{\rm kep}^{2} = K \frac{ r_{x}^{2} } { (r_{x}^{2} + \beta^{2}r_{x}^{2\alpha})^{3/2} } \, ,
\end{equation}

\noindent where $v_{\rm kep}$ is the local Keplerian velocity and
$K=GM_{\rm BH}$, where $M_{\rm BH}$ is the mass of the black hole.
           In the limiting case of a geometrically thin disc,
            (i.e. $\beta=0$), the velocity field reduces to that
            expected for simple planar Keplerian orbits.  Significant
            radial motion, e.g., bulk radial motion or scale height
            dependent turbulence (Collin et al. 2006), may be included
            by introducing an azimuthal perturbation to the velocity
            field (see e.g. Goad, Korista and Ruff (2012), their
            equation 4.)  We adopt $\alpha=2$ and a time-delay at the
          outer radius $\tau(r = R_{\rm out}) = (r - H)/c = 50$~days,
          chosen to match the dust-delay reported for the Seyfert 1
          galaxy NGC~5548 (Suganuma et al. 2004, 2006; Koshida et
            al. 2014), yielding $\beta=1/150$.  We assume a black
          hole mass of $10^{8}$M$_{\odot}$, similar to the
          best-estimate of MBH for this source derived from
          reverberation mapping experiments (Peterson and Wandel
          2000). For the continuum normalisation appropriate for
          NGC~5548 the fiducial BLR geometry spans a radial distance
          of between 1.14--100 light-days.  Here we set the BLR
            inner radius to 200 gravitational radii. However,
            differences of a factor a few in $R_{in}$, arising from
            differences in the adopted value of $M_{BH}$ can be
            tolerated because in general (i) the radial surface
            emissivity distribution $F(r)$ decreases at the smallest
            BLR radii for most lines, and (ii) the surface area of BLR
            clouds at small BLR radii is small. Taken together the
            contribution to the total line emission of gas located at
            small BLR radii is small (see \S5.1).  The BLR outer
            radius is here set by the distance at which robust
            graphite grains can survive.  For the bowl inclination,
          we adopt $i=30$ degrees, close to the value reported for
          NGC~5548 by Pancoast et al. (2014b), a value considered to
          be typical of the expected inclination of type~{\sc i}
          objects.

\subsection{Radial surface emissivity distributions}\label{surface}

\begin{figure}
\resizebox{\hsize}{!}{\includegraphics[angle=0,width=8cm]{fig1a.ps}}
{{\bf Figure~1a.} Contours of predicted $\log$(EW) for broad
  H$\beta$ referenced to the incident continuum flux at
  $\lambda$1215\AA\, for full source coverage at each point in the
  grid, as a function of gas hydrogen density $n_{\rm H}$ and
  hydrogen-ionizing photon flux $\Phi_{\rm H}$. The total hydrogen column
  density for each cloud in the grid is $\log N_{\rm
    H}$(cm$^{-2}$)$=23$. The smallest contour corresponds to 0.1\AA\;,
  each solid line represents 1 decade, and dotted lines represent 0.2
  dex intervals. Contours of $\log(EW) < -1$ are not plotted (upper
  left quadrant). The solid star marks the old "standard BLR"
  parameters, the solid triangle the peak EW. The dot-dashed red lines
  indicate the range in $\log Uc \equiv \log (\Phi_{\rm H}/n_{\rm H})$
  (diagonal lines) and $\log n_{H}$ (vertical lines) used to compute
  the radial surface emissivity distribution (Figure~1b, upper
  panel).}
\label{fig1a}
\end{figure}

\addtocounter{figure}{-1}

\begin{figure}
\resizebox{\hsize}{!}{\includegraphics[angle=0,width=8cm]{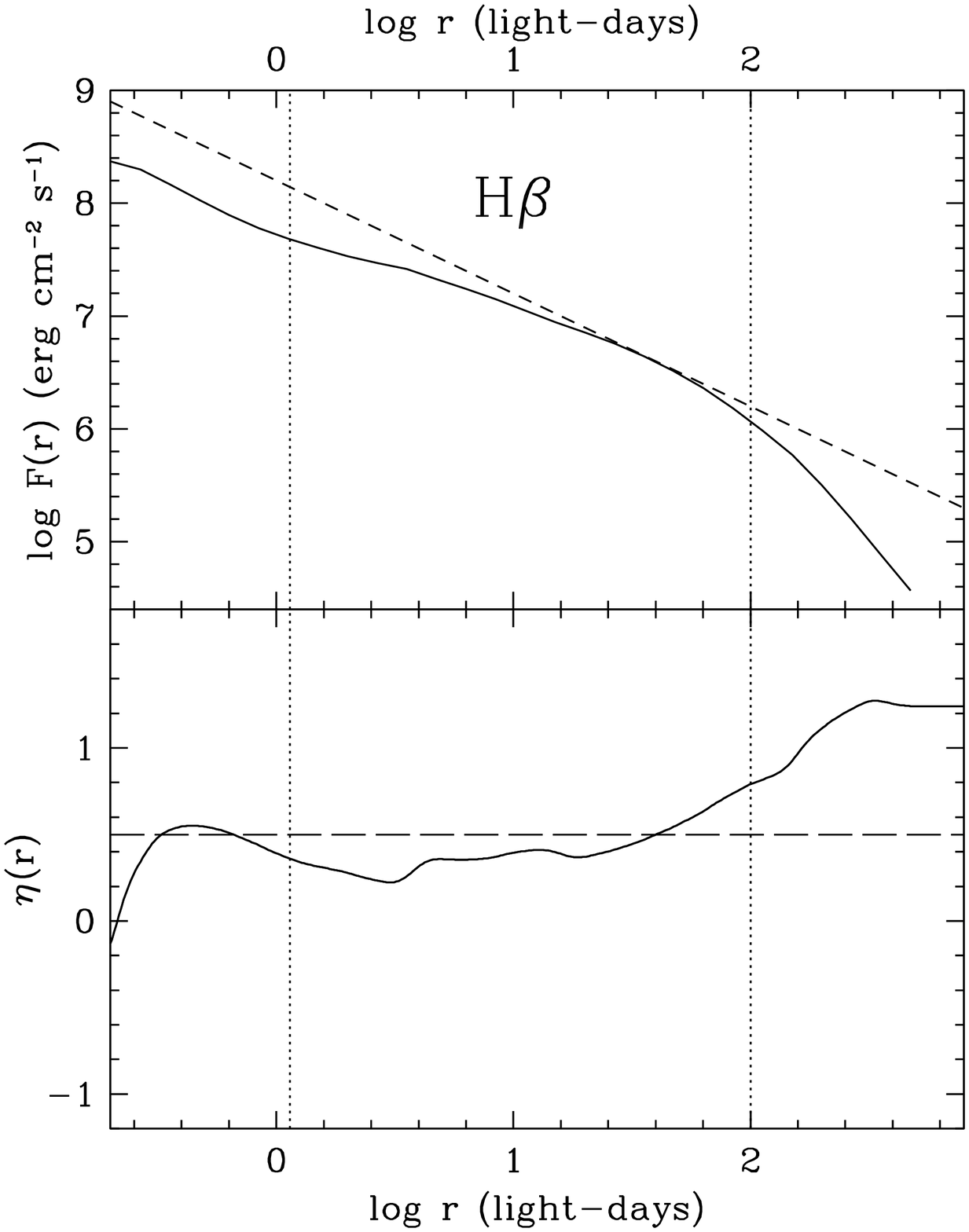}}
{{\bf Figure~1b.} Upper panel -- an LOC model prediction of the radial
  surface emissivity distribution $F(r)$ for broad H$\beta$ (solid
  black line, see text for details). For purposes of comparison, the
  dashed black line indicates a power-law radial surface emissivity
  distribution $F(r) \propto r^{\gamma}$, with $\gamma=-1.0$, for
  which $\eta(r)=-(\gamma/2)=0.5\; \; \forall r$. This power-law
  radial emissivity distribution is tangent to the $F(r)$ distribution for broad
  H$\beta$ at a distance of $\approx 30$~light-days. Lower panel --
  the corresponding radial responsivity distributions. The local
  responsivity for broad H$\beta$ is smaller than $\eta(r)=0.5$ for
  BLR radii $< 30$~light-days, and larger than $\eta(r)=0.5$
  for radii $> 30$~light-days. The dotted vertical lines indicate the location of the inner and
  outer boundaries of the fiducial BLR.}
\label{plot_eta_hb}
\end{figure}

In paper~{\sc i} we adopted simple power-law radial surface emissivity
distributions $F(r)\propto r^{\gamma}$, with power-law indices
$\gamma$ chosen to broadly match the range predicted by
photoionization model calculations ($-2 \leq \gamma \leq -1$). Under
these conditions, the radial emission-line responsivity $\eta(r)$ is
constant both spatially and temporally $\eta(r) = -(\gamma/2) =
constant$ (we assume here, and throughout, that the incident ionizing
continuum flux scales as $r^{-2}$).  In this contribution we use the
radial surface emissivity distributions for individual lines as
described in Goad, Korista and Ruff (2012).

In brief, we computed a grid of photoionization models assuming simple
constant density slabs of gas with fixed constant total hydrogen
column density $\log N_{\rm H}$(cm$^{-2}$)$ = 23$, solar abundances,
and each with a direct view of the ionizing continuum source. Unless
otherwise stated, here photoionization model calculations were
performed with Cloudy, version C08.00 (Ferland, Korista and Verner
1997; Ferland et al. 1998), adopting a modified version of the
Matthews and Ferland (1987) generic AGN continuum spectral energy
distribution (see Goad, Korista, and Ruff 2012 for details). We note
that while this incident continuum SED is likely softer (smaller
X-ray/UV power ratio) than is expected in NGC~5548, these differences
have relatively small effects on the predicted hydrogen line spectrum
(Korista et al. 1997). Here, the incident continuum has been scaled to
match an estimate for the mean ionizing continuum luminosity of
NGC~5548, $\log L_{\rm ion}$ (erg~s$^{-1}$)=44.14, based on the Galactic
extinction corrected mean UV flux at 1350\AA\, measured during the
1993 HST monitoring campaign (Korista et al. 1995)
\footnote{In this grid of photoionization models, the maximum
  principal quantum number, n, with angular momentum resolved levels
  was increased above its default value in the model atoms of H~{\sc
    i}, He~{\sc i}, and He~{\sc ii}: to n=18, 15, and 15,
  respectively. Those levels above were l-averaged. This was done to
  utilise the improved accuracy of the more sophisticated model atoms
  appearing in Cloudy version C08. We caution that the default model
  atoms are unlikely to predict accurate H, He spectra from
  photoionized gas with physical conditions expected in the BLR.}.
For the adopted ionizing continuum shape and normalisation, a hydrogen
ionizing photon flux $\log \Phi_{\rm H}$ (photon~cm$^{-2}$~s$^{-1}$) =
20.0 corresponds to a continuum source distance of $R =
15(D_{L}/77.6$~Mpc)~light-days.  The full grid spans seven decades in gas
hydrogen number density $n_{\rm H}$ and hydrogen ionizing photon flux
$\Phi_{\rm H}$, $7 < \log n_{\rm H}$~(cm$^{-3}$)~$ < 14$, and $17 <
\log \Phi_{\rm H}$~(cm$^{-2}$~s$^{-1}$)~$ < 24$, stepped in 0.25
decade intervals in each dimension (see e.g. Korista et al. 1997).

Figure~1a indicates the logarithm of the equivalent width (EW) of
H$\beta$ (hereafter EW(H$\beta$)) referenced to the incident continuum
flux at $\lambda$1215\AA\ , as functions of $\log n_{\rm H}$ and
$\log \Phi_{\rm H}$. Solid lines represent a decade in EW starting
from $\log(EW)= -1$ at the upper left to $\log(EW)=2$ at the lower
right. The dotted lines indicate 0.2 dex intervals. For reference, the
old `standard BLR' parameters (Davidson and Netzer 1979) are marked by
the solid star, while the peak EW for this emission line is marked by
the solid triangle. The red (dot-dashed) lines indicate the boundaries
in $\log Uc$ (diagonal lines) and $\log n_{\rm H}$ (vertical lines)
used when calculating the radial surface emissivity distribution for
broad H$\beta$ (Figure~1b, upper panel), described below. The upper
bound in $\log Uc \approx 12$ is representative of conditions at which
clouds become thermally unstable, and so unlikely to exist stably.  We
also imagine that the total pressure in the environment depends on the
depth within the gravitational potential well of the supermassive
black hole, and thus we've also set a lower bound in $\log Uc$.  See
Korista and Goad (2000, 2004).


The line EW is a measure of the efficiency by which ionizing continuum
photons are converted into emission line photons.  The rapid decline
in the line EW(H$\beta$) near the diagonal line $\log \Phi_{\rm H}
\approx \log n_{\rm H} + 10.7$ is due to the hydrogen in the fixed
column density slabs becoming fully ionized.  Gas near the Compton
temperature can be found in the upper left corner of Figure~1a, and
the contours of these insignificant EWs are not plotted for
clarity. For H$\beta$ as for the other hydrogen and helium optical
recombination lines the EW increases with increasing density, a
consequence of increasing contributions from collisional excitation
(Ferland and Netzer 1979). The general decline in EW(H$\beta$) in the
direction of increasing incident ionizing photon flux for $\log
\Phi_{\rm H}$~(photons~s$^{-1}$~cm$^{-2}$) $>$ 18 is a consequence of
increasing line optical depth in this direction, and then eventually
increasing photoionizations from excited states. This steady decline
in EW(H$\beta$) with increasing incident ionizing photon fluxes, noted
above, indicates clouds with responsivities $\eta < 1$ (see equation
2), for this emission line.
Where the EW contours are sparse or well-separated with respect to
changes in the incident ionizing photon flux, the line EW$\approx
constant$, and thus these clouds have responsivities $\eta \approx 1$.
Increasing values in EW with respect to increasing values in incident
ionizing photon flux indicate clouds with responsivities $\eta >
1$. This is predicted in H$\beta$ at only the smallest incident
ionizing continuum fluxes (see Figure~1a).

Average radial surface emissivity distributions for individual lines
are generated by summing over the gas density distribution ($8 < \log
n_{\rm H}$ (cm$^{-3}$)~$ < 12$) and $\log Uc = \log \Phi_{\rm H} -
\log n_{\rm H}$, as described in Korista and Goad (2000).  We use the
standard LOC gas density distribution weighting function $g(n_{\rm
  H})\propto n_{\rm H}^{-1}$ described in Korista and Goad (2000; see
also Baldwin et al. 1995; Bottorff et al. 2002), and which roughly
matches the gas density distribution of fragmenting BLR clouds
resulting from magnetohydrodynamic instabilities (Krause et al. 2012).
The chosen ranges in $\log n_{H}$, $\log Uc$, as indicated by the red
dashed-lines in Figure~1a, are nearly identical to those adopted in
Korista and Goad (2000, 2001, 2004). In order to investigate the
continuum-driven variability of broad H$\beta$, we compute the radial
surface emissivity distribution with respect to incident photon fluxes
well below that required to sublimate robust graphite grains at the
incident face of the cloud, and likewise well inside the fiducial
inner radius of the BLR. As in previous work, we assume an open
geometry. That is, we do not address the effect of cloud--cloud
shadowing of the incident continuum photons nor the partially
transmitted and diffuse continuum and emission line photons, nor their
interaction with the rest of the cloud population on their passage
through the BLR.

The model radial surface emissivity distribution $F(r)$ for H$\beta$
is shown in Figure~1b (upper panel, solid black
line). Also shown is the corresponding radial responsivity
distribution (lower panel, solid black line). Over the radial extent
of the fiducial BLR geometry (indicated by the vertical dashed lines),
$F(r)$ may be approximated by a broken power-law with slope $\gamma
\approx -0.7$ for $r<25$~light-days, breaking to a steeper slope,
$\gamma > -2$ for larger radii. This corresponds to a range in radial
responsivity $\eta(r)$ of $0.35 < \eta(r) < 1.0$ for this line.  The
effective responsivity for a particular BLR geometry can be
determined using a weighted average of the radially dependent
responsivities of individual clouds and will, in the absence of significant
geometric dilution and windowing effects, lie somewhere within this
range.  Significantly, this range in $\eta(r)$ values is similar to the
range in measured responsivity $0.4 < \eta_{\rm eff} < 1.0$ for this line
obtained from an analysis of each of the 13 seasons of monitoring data
for NGC~5548 (Goad et al. 2004), though the latter is referenced to
the optical continuum and not the UV continuum variations.  The dashed
lines in Figure~1b represent a power-law radial surface
emissivity distribution $F(r)$, with power-law index $\gamma=-1$
(upper panel), which equates to a radial responsivity distribution
$\eta(r) = -(\gamma/2) = 0.5$ (lower panel) and is for comparison
purposes only.

\subsection{The driving continuum light-curve}\label{continuum}
  
To model the broad emission-line variations in NGC~5548 we first
require an appropriate driving ionizing continuum light-curve. Here we
generate what we refer to as a "mock" ionizing continuum light-curve
for NGC~5548 using as a template the host-galaxy subtracted variable
optical continuum light-curve from the $13+$ years of ground-based
monitoring of NGC~5548 by the AGN~Watch collaboration (Peterson et
al. 2002; Peterson et al. 2013). To remove the non-variable host
galaxy contribution to the optical continuum, we use the updated value
from Bentz et~al. 2006, derived from HST images and scaled to the
appropriate ground-based aperture, and which is approximately 10\%
larger (3.75 c.f. 3.37 $\times
10^{-15}$~erg~cm$^{-2}$~s$^{-1}$\AA$^{-1}$) than that used by
Romanishin et al. 1995. We then scale the galaxy subtracted optical
continuum according to the best estimate of the measured relationship
between the UV continuum and optical continuum variability (
$F_{\lambda 5100} \propto F_{\lambda 1350}^{\beta}$ ), with
$\beta=0.84 \pm 0.05$ (Bentz et al. 2007) thereby generating a proxy
for the driving ionizing continuum light-curve for this
source\footnote{Upon completion of this work we discovered that the
  quoted value of $\beta$ had not been corrected for extinction within
  our Galaxy (Bentz 2014, private communication). The de-reddened
  value of $\beta \approx 0.75$ results in a 20\% increase in the
  continuum fluctuation amplitude. Significantly, when combined with
  the measured relation between H$\beta$ delay and optical continuum
  luminosity $\tau({\rm H}\beta) \propto L_{\rm opt}^{0.66 \pm 0.13}$,
  the size--luminosity relation becomes $r_{\rm BLR} \propto L_{\rm
    uv}^{0.495}$ in line with photoionization model predictions. We
  note that this does not substantially alter the findings presented
  in this work.}. While modifying the optical continuum in this way
provides a reasonable approximation to the amplitude of the UV and so
presumably the ionizing continuum variations, it does not reflect the
true characteristic timescale of the UV continuum, since the optical
continuum is generated at larger and over a broader range in disc
radii than the UV continuum. As the designated driver, this
  continuum is by definition at zero delay with respect to longer
  wavelength continuum variations and broad emission-line variations.
Additionally, as reported in Korista and Goad (2001), the optical
continuum is contaminated by a more slowly varying diffuse continuum
component arising from BLR clouds and which may represent as much as
20\% of the optical continuum flux at 5100\AA\, for NGC~5548. Finally,
the limited studies available for the EUV continuum in this source
suggests that the variable ionizing EUV continuum displays even larger
amplitude variability (Marshall et~al. 1997; Chiang et al. 2000) than
that used here.

This light-curve is then re-sampled onto a regular 1-day grid by
interpolating between bracketing points with uncertainties estimated
using a structure function analysis (see e.g. Goad, Korista and Knigge
2004; Goad and Korista 2014, paper~{\sc i}).  Using the mock ionizing
continuum we drive the fiducial BLR geometry for a single broad
emission-line H$\beta$, allowing us to compare model emission-line
variations with the most extensively studied broad emission-line for
which the best possible data is available.

\section{Simulations}\label{sims}

\subsection{Reverberation mapping :  Forward modelling}

Successful forward models of the broad emission-line region must not
only match the observed emission-line variations (amplitude and delay)
about their mean level, but in addition, {\em should also satisfy the
  necessary energy requirements for that line\/} (Korista and Goad
2000; Kaspi and Netzer 1999; Horne, Korista and Goad 2003).  While it
is not the intent of this work to model the broad emission lines and
their variability in NGC~5548 in detail, we have checked that the
fiducial model geometry (integrated out to 100 light days) and LOC
model emissivities predict a {\em sum\/} in the luminosities of
Ly$\alpha$ $\lambda$ 1216 \AA\ and C~{\sc iv} $\lambda$ 1549\AA\ that
match the average value from the 1993 HST monitoring campaign (Korista
et al. 1995; Korista and Goad 2000).

\subsection{The importance of F(r) to the emission line response}

First, we investigate how differences in the radial surface emissivity
distribution impact upon the model emission-line light-curves. We
generate emission-line light-curves using power-law radial surface
emissivity distributions ($F(r) \propto r^{\gamma}$), and power-law
indices $\gamma=-2$ and $\gamma=-1$, equivalent to radial responsivity
distributions of $\eta(r)=1.0$ and $0.5\; \forall\/ r$ respectively,
and a physically motivated model for the radial surface emissivity
distribution for broad H$\beta$ (Figure~1b, upper panel) constructed
for an LOC model of the BLR in NGC~5548, and for which $0.35 < \eta(r)
< 1.0$.  These we compare with the observed 13~yr H$\beta$
emission-line light-curve for NGC~5548. For the power-law radial
surface emissivity distributions we assume isotropic line
emission. For the physically motivated model, we adopt a line
radiation pattern that approximates the phases of the moon (e.g.
O'Brien, Goad and Gondhalekar 1994), with the inward fraction equal to
80\% of the total (e.g. Ferland et al. 1992; O'Brien, Goad, and
Gondhalekar 1994). For a bowl-shaped geometry, differences in the
adopted form of the emission-line anisotropy have little effect on the
amplitude of the emission-line response (Goad and Korista 2014, their
figure 17), and have a similarly small effect on the emission-line
delays, for the range in line-of-sight inclination expected for type 1
AGN.

Model emission-line light-curves are generated by
driving the fiducial BLR model with our mock ionizing continuum,
assuming a fully non-linear response in the line\footnote{In practice, we
  calculate the local radial surface emissivity distribution $\eta(r)$
  at the current epoch from the steady-state radial surface emissivity
  curve $F(r)$ for that line but shifted according to the continuum
  level at that epoch (i.e. $\eta(r) = \eta(r, L(t))$). The
  emission-line light-curve is then determined by summing over the
  radial surface emissivity distributions (at each epoch), scaled
  according to the radial covering fraction dependence, and with an
  appropriate correction for the light-crossing time from the
  continuum source to the line-emitting region.} for a BLR with fixed
inner and outer boundaries.

The results of this study are shown in Figure~\ref{plot_hb_pow_large}.
The mock driving ionizing continuum light-curve (black points) and the
narrow emission-line subtracted broad H$\beta$ light-curve (red
points) as determined using the latest values for the variable narrow
emission-line contribution in this source (Peterson et al. 2013) are
shown in panel~(i).  Here and elsewhere, the continuum and
emission-line light-curves have been plotted after first normalising
to their respective mean values, as calculated from epoch 100
onwards\footnote{We could have instead normalised the model
  emission-line light-curves to their respective steady-state values,
  and compared these to the observed continuum and emission-line
  light-curves averaged over the full 13~yr campaign. We have verified
  that these differences in the normalisation affect the light-curves
  shown here at less than the few percent level.}.  This is
  sufficient to allow the whole of the fiducial BLR to respond. For
  the fiducial bowl-shaped model, the maximum delay at the outer
  radius of 100 light-days when viewed face-on is only 50 days because
  gas at larger BLR radii is elevated relative to the mid-plane of a
  face-on disc and thus lies closer to the observer line of sight
  (i.e., the bowl has a radially dependent scale height, see equation
  3). When viewed at an inclination of 30 degrees, the maximum delay
  at the outer radius increases to 100 days (see also Figures~1, 4
  of Goad, Korista and Ruff (2012)).
In order to make a quantitative comparison, we first add
noise to each of the model emission-line light-curves, by adding a
random Gaussian deviate to each point, with dispersion $\sigma$ equal
to 1\% of the flux, and assigning an error bar in a similar
fashion. We then compute the continuum--emission-line
cross-correlation function (hereafter, CCF), from which we measure the
peak delay (or lag) and the centroid (equivalent to the
luminosity-weighted radius of the BLR). CCFs are constructed using the
implementation championed by Gaskell and Peterson (1987),
interpolating on both light-curves. The CCF centroid is measured over
the range in delays for which the cross-correlation coefficient
exceeds a 0.8 of its peak value.  In addition, we measure the average
emission-line responsivity for the full 13~yr light-curve using the
ratio $d\log F_{\rm line}/d \log F_{\rm cont}$ after first correcting
for the mean continuum--emission-line delay (see Goad, Korista and
Knigge (2004), for details).  The delays and responsivities reported
in Table~2 for these and all other observed and simulated light-curves
have been measured relative to the mock driving ionizing continuum.

\addtocounter{figure}{2}

\onecolumn
\begin{figure}
\resizebox{\hsize}{!}{\includegraphics[angle=270,width=8cm]{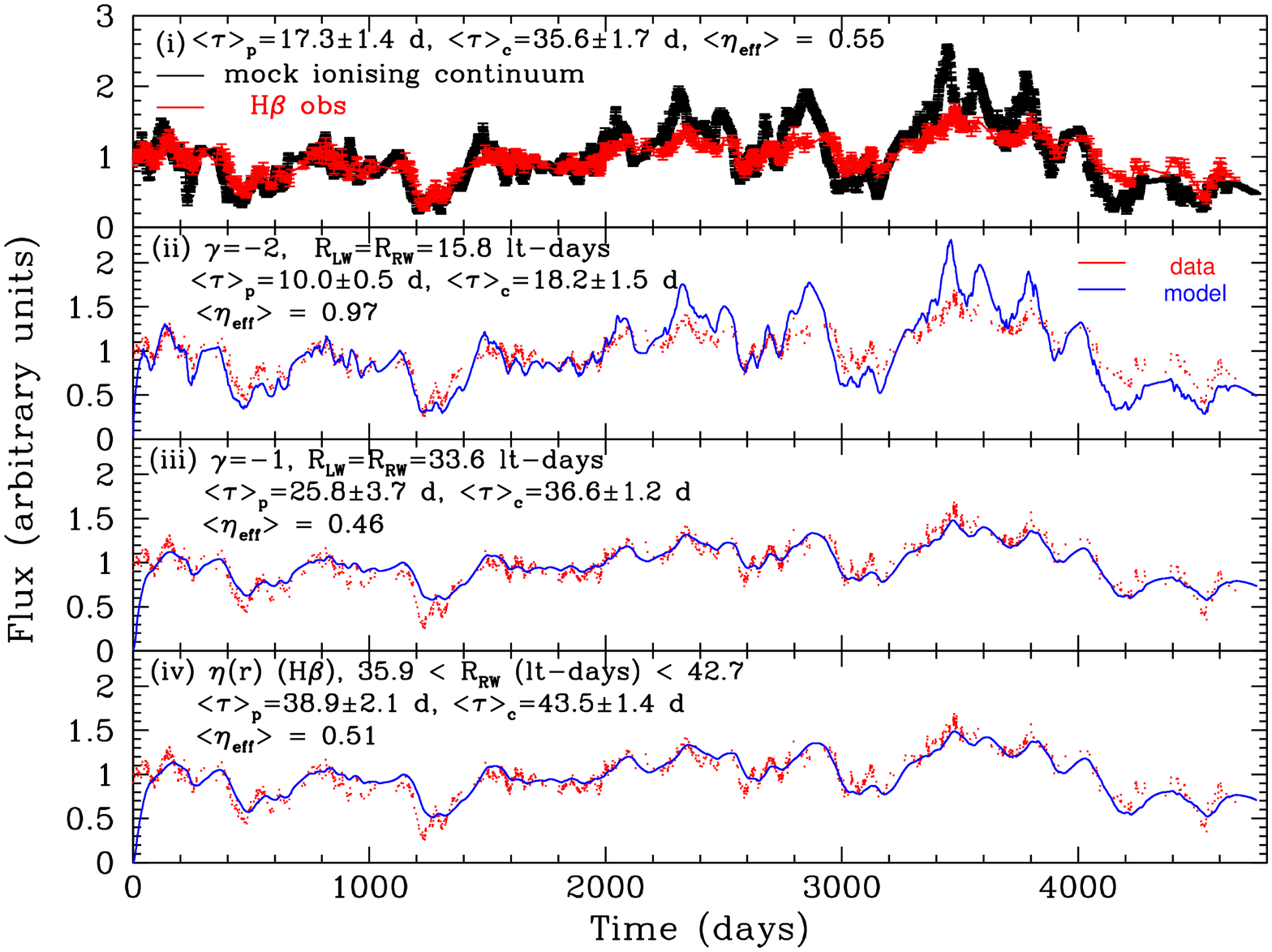}}
\caption{Panel (i) -- the mock ionizing continuum flux light-curve
  (black points) and updated narrow-line subtracted broad H$\beta$
  flux light-curve (red points). Panels 2 and 3 -- the observed broad
  H$\beta$ flux light-curve (red points) together with model
  emission-line flux light-curves (solid blue lines) for power-law
  radial surface emissivity distributions $F(r) \propto r^{\gamma}$,
  with $\gamma=-2$, panel (ii), and $\gamma=-1$, panel (iii). Panel
  (iv) -- the observed broad H$\beta$ flux light-curve (red points),
  together with the model emission-line flux light-curve, assuming a
  radial surface emissivity distribution from the fiducial LOC
  model. All light curves have their mean values calculated from day
  100 onwards normalised to 1.}
\label{plot_hb_pow_large}
\end{figure}

\twocolumn

Figure~\ref{plot_hb_pow_large} panels (i)--(iii) illustrate a number
of key points.  The measured delay (CCF centroid, lag) and
time-averaged responsivity $<\eta_{\rm eff}>$ for H$\beta$ for the
full 13~yr ground-based observing campaign of NGC~5548 {\em relative
  to the mock ionizing continuum\/} are $35.6 \pm 1.7$~days (CCF
centroid), lag $17.3 \pm 1.4$~days (CCF peak) and $<\eta_{\rm
  eff}>=0.55$ (panel (i)).  These values are to be compared with
measured delays of $18.2 \pm 1.5$~days and $36.6 \pm 1.2$~days (CCF
centroid), $10.0 \pm 0.5$~days and $25.8 \pm 3.7$~days (CCF Peak) and
measured responsivities $<\eta_{\rm eff}>\approx0.97$ and $\approx
0.46$ for light-curves generated using power-law radial surface
emissivities with power-law indices $\gamma=-2$ (panel (ii)) and
$\gamma=-1$ (panel (iii)) respectively.  

Power-law radial surface emissivity distributions have by construction
$\eta(r)=constant$; \; both spatially and temporally, and consequently
the responsivity-weighted radius $R_{\rm RW}$ is independent of
continuum level, i.e., these models cannot breathe. These models also
show little evidence for geometric dilution (on average) of the emission line
response. That is, the measured time-averaged responsivities $<\eta_{\rm eff}>=0.97$
and 0.46, are 97\% and 92\% of their expected values, while the delays
determined from the centroid of the CCF are close in value to the
responsivity-weighted radii ($R_{\rm RW}$).  This is to be compared
with the significant geometric dilution ($\approx 20$\%) exhibited by
the same geometry when viewed face-on (for which geometric dilution is
minimised) for the same power-law radial surface emissivity
distributions driven by fake continuum light curves with
characteristic variability timescale $T_{\rm char}=40$~days (e.g.,
Figure~14, Goad and Korista 2014, paper~{\sc i}).  We infer from this
that the mock ionizing continuum variability light curve we've adopted
here to drive our models has a $T_{\rm char}$ substantially longer
than 40 days. Whether this is the case or not in NGC~5548 is unclear,
since while the model continuum light curve likely has a more
realistic variability amplitude, it remains in essence a time-blurred
version of the true driving continuum (\S2.2).

The second point of note is that while a steep radial surface
emissivity distribution ($F(r) \propto r^{-2}$) is a better match to
the observed emission-line variations at the start of the 13
yr~campaign (panel (ii), epochs 0--1200), it displays variability over
and above that which is observed at later times (e.g., epochs 2000
onwards). We note that even at the start of the campaign, the
variability amplitude is too large for this model, and the delay too
small on average when compared to the observations, suggesting that
the power-law index for $F(r)$ is flatter than $-2$, so that $\eta(r)$
is in general less than 1. Figure~\ref{plot_hb_pow_large} panel (iii)
suggests that a shallower radial surface emissivity distribution
provides a better match to the observed broad H$\beta$ variations
during high continuum states (e.g. epochs 2000--4000), but is
generally a poorer match during low-continuum states ($\approx
400$--$1400$).  That the observed H$\beta$ emission line behaviour
appears to fall between these bounds in responsivity (0.5--1.0) is
likely a manifestation of the luminosity-dependent behaviour of the
emission line responsivity (Korista and Goad 2004).

We emphasise that no attempt is made to fit the data. Rather, the
models we present serve to highlight the observed behaviour of the
emission-line light-curves about their respective mean levels, and to
illustrate the key physics important in determining not only the
emission-line luminosity, but also its response amplitude and delay,
for a particular choice of BLR geometry\footnote{Currently, dynamical
  models of the BLR (Pancoast et al. 2012, 2014a,b) do not account for
  the spatially and time-variable responsivity of the line-emitting
  gas, assuming instead a constant value for $\eta(r,t)=1$ (i.e., a
  strictly linear response).  These models also do not predict the
  emission line power from the model geometry. It seems unavoidable,
  however, that the inferred system geometry, emission line
  luminosity, emission line responsivity and delay are inextricably
  tied together (and supported by the well known BLR
  radius--luminosity relation, $R_{\rm BLR} \propto L_{\rm
    UV}^{1/2}$).  The present work reinforces the importance of
  incorporating the physical properties of the line-emitting gas into
  those forward modelling techniques employed to recover the spatial
  distribution and kinematics of the line-emitting gas from
  reverberation mapping data.}.  Since a steeper radial surface
emissivity distribution appears a better match to the observed
variability behaviour of broad H$\beta$ at early times, then in the
context of our chosen BLR geometry, this suggests that at the start of
the campaign the BLR is both more compact (which for a given geometry
implies a smaller responsivity-weighted radius, and not necessarily a
smaller BLR) and locally has a larger responsivity.  Conversely,
during the middle and latter parts of the 13~yr campaign, H$\beta$
shows a weaker amplitude response, due to a smaller local
emission-line responsivity and increased responsivity-weighted
radius. During the latter half of the campaign the continuum level is
larger than at the start of the campaign. Thus the response amplitude
in H$\beta$ appears to be anti-correlated with continuum level -- a
behaviour identified in greater detail by Goad, Korista and Knigge
(2004), and in qualitative agreement with the predictions of
photoionization models (Korista and Goad 2004).

\subsection{A physical model for $F(r)$}
As noted by Korista and Goad (2000, 2004), Goad, Korista and Ruff
(2012), and Goad and Korista (2014), the radial surface emissivity
distributions of the broad emission lines are in general a poor
approximation to a simple powerlaw. This has the important consequence
that even for a BLR with static inner and outer boundaries, the BLR
may still breathe (e.g. Goad and Korista 2014).  In Figure~2 panel
(iv), we show the model H$\beta$ emission-line light-curve generated
using the radial surface emissivity distribution $F(r)$ from Figure~1b
(upper panel), generated for the fiducial LOC model of NGC~5548.  When
compared to power-law radial surface emissivity distributions, this
model exhibits a number of promising characteristics.

First, the fiducial model's responsivity averaged over the full 13~yr
campaign, $<\eta_{\rm eff}>=0.51$ (panel (iv) of Figure~2), is a good
approximation to the measured time-averaged value for this line
$<\eta_{\rm eff}>=0.55$ when referenced to our mock ionizing
continuum. Since geometric dilution is small for the adopted driving
continuum and BLR geometry (\S3.2), then this radial responsivity for
broad H$\beta$ determined from a physically motivated model provides a
better match to the time-averaged responsivity of this line compared
to those predicted by simple power-law description of $F(r)$ (panels
(ii) and (iii) of Figure~2, \S3.2)

Second, for a radial surface emissivity distribution $F(r)$ that
deviates significantly from a simple power-law (Figure~1b, upper
panel, solid line versus dashed line), {\em the responsivity-weighted
  radius $R_{\rm RW}$ and measured line responsivity $\eta_{\rm eff}$
  will vary with continuum level\/}. Since $F(r)$ steepens at larger BLR
radii (i.e., towards lower ionizing continuum fluxes, Figure~1b, upper
panel), the effective responsivity $\eta_{\rm eff}$ will increase
during low continuum states (Figure~1b, lower panel), due to the
larger on average emission-line responsivity and smaller
responsivity-weighted radii.

Thus a physically motivated description of the radial surface
emissivity distribution for broad H$\beta$ produces a BLR model which
not only breathes, but does so in the correct sense: in low continuum
states the BLR responds on shorter timescales and with larger
amplitude than in high continuum states, as is observed (Goad, Korista
and Knigge 2004).  However, while the responsivity weighted radius
correlates with continuum flux, its range is rather modest $35.9 <
R_{\rm RW}$ (light-days)$ < 42.7$. Similarly, if we divide the
continuum and emission-line light-curves shown in panel (iv) into
contiguous segments, each spanning $\approx$ 1000 days, $\eta_{\rm
  eff}$ is found to vary from $0.47 < \eta_{\rm eff} < 0.75$ and
importantly anti-correlates with continuum level. Both behaviours are
clear signatures of breathing (see Gilbert and Peterson 2003; Goad,
Korista and Knigge 2004; Cackett and Horne 2006; Kilerci Eser et al. 2015).

However, the fiducial LOC model does not match the deep excursions
exhibited by broad H$\beta$ during low continuum states (e.g. epochs
1200, 4200), nor in detail the short timescale variations.  This
largely arises because the fiducial BLR is too large, with a measured
delay of $43.5\pm 1.4$~d (CCF centroid), $38.9 \pm 2.1$~d (CCF peak),
$\approx$ 20\% larger (CCF centroid) than that measured for the 13~yr
campaign when referenced to our mock driving ionizing continuum. Thus
while this model BLR can breathe, and provides a reasonable match to
the observed emission-line variations during high continuum states
(e.g. epochs 1500--4000), it is still too large on average and
responds too weakly during low-continuum states, i.e., within the
confines of the BLR boundaries, the adjustments in the local radial
surface emissivity in response to continuum variations are not large
enough to significantly modify the mean response timescale and
amplitude of response in the line. Additionally, the fiducial LOC
model and BLR geometry predicts a mean luminosity $\log L$(H$\beta$)
(erg~s$^{-1}$) = 41.497, $\approx$ a factor 2 smaller than the
measured 13~yr time-averaged, narrow-line subtracted luminosity for
broad H$\beta$, once corrected for Milky Way extinction, $\log
L$(H$\beta$) (erg~s$^{-1}$) = 41.756 (see Table~1).

In paper~{\sc i} we showed that the measured emission-line response
amplitude and delay, for a particular choice of geometry, depend upon
(i) the local responsivity in the line-emitting gas, (ii) the
monitoring campaign duration and (less so) sampling rate, and (iii)
the amplitude and characteristic timescale $T_{\rm char}$ of the
driving ionizing continuum relative to the maximum delay $\tau_{max}$
at the BLR outer radius for a range of plausible geometries given
observer line of sight orientation.  In the next section we explore
the connections between the emission-line luminosity, responsivity,
and delay to continuum variations, as well as effects which alter
these key quantities without altering the underlying BLR geometry.

\section{The link between emission-line luminosity, characteristic size 
and responsivity}

For an assumed BLR geometry, the emission-line luminosity is
determined by integrating over the radial surface emissivity
distribution $F(r)$, weighted according to the BLR cloud distribution
and covering fraction. Differences in $F(r)$ for the same geometry
will alter : the integrated emission-line luminosity (the energy of
the system), the measured continuum--emission-line delays, and the
effective emission-line responsivity. For example, if we consider the
power-law radial surface emissivity distributions, illustrated in
Figure~2, panels (ii) and (iii), a steeper $F(r)$ (i.e. $\gamma$ more
negative) results in a smaller emission-line luminosity (assuming that
$F(r)$ is normalised to the same value at the BLR inner radius),
smaller delays and characteristic sizes, and a larger
responsivity. Thus, these quantities can not be treated in isolation
but are instead inextricably tied together. Indeed all of these
quantities are contained within the continuum--emission-line transfer
function $\Psi(\tau)$ (Blandford and McKee 1982), the function which
maps the continuum light-curve on to the integrated emission-line
light-curve.

We note that the transfer function $\Psi(\tau)$ and the response
function $\Psi^{\prime}(\tau)$ have often been used interchangeably in
the literature and as a consequence have been a source of much
confusion.  However, they are not the same.  Here we distinguish
between the emission-line transfer function $\Psi(\tau)$, which
contains the total light from the BLR, and the emission-line response
function $\Psi^{\prime}(\tau)$, which deals with only the variable
part of the line emission. The latter is recovered from a linearised
version of the transfer equation, where constant (or slowly varying)
components from both the line and continuum are confined to the
background; i.e. $\Psi^{\prime}(\tau)$ represents the partial
derivative of the line with respect to continuum variations. Weakly
responding gas present in the BLR will contribute to the total light
in a particular emission line, but less so to its variable light, and
so may not be recovered in the response function.  Using this
distinction, transfer functions and response functions will generally
not look the same (they do have the same shape for a powerlaw
emissivity distribution because in this case the line responsivity is
constant everywhere).

While a different geometrical configuration for the BLR gas could act
to enhance the emission-line responsivity, by reducing the delays and
thereby increasing the coherence of the emission-line response, {\em
  in the absence of geometric dilution, it is the line responsivity
  that determines the amplitude of the emission-line response to
  continuum variations\/}.  Furthermore, one cannot simply change the
geometry, without affecting: (i) the measured continuum--emission-line
delays, (ii) the amplitude of the emission-line response (the
responsivity) and (iii) importantly the luminosity for a particular
line. Attention to the geometry has mostly focused on the
emission-line delays (or lags), yet all three are intimately
connected.  

We illustrate these connections with two examples.  First, reducing the
BLR outer boundary incurs a significant penalty in terms of the total
emitted power of a particular line, due to the significant loss in
surface area which would otherwise be available at large BLR radii
(e.g. compare the fiducial model in Table~1 with the truncated model
in which $R_{\rm out} $ is set to 50 light-days). This statement holds
for any BLR geometry in which the covering fraction dependence is not
a steep function of radius. This is especially relevant to the
hydrogen recombination lines, which tend to be more emissive as well
as more responsive at large BLR radii. Thus to accommodate a smaller
BLR outer boundary, a means must be found for boosting the line
emissivity and responsivity at smaller radii, such that the energy
requirements for a particular line are also satisfied.  Secondly, we
could also reduce the discrepancy in the fiducial models predicted
luminosity by allowing for a larger BLR outer boundary. However, this
would come at the expense of still longer continuum--emission-line
delays.  Attempts to recover the broad emission-line geometry and
kinematics which do not account for the power in the continuum and
lines should be treated with caution.

\subsection{Parameters governing physical conditions that affect H$\beta$ 
luminosity, characteristic size, and responsivity}

Here we explore those effects which act to alter the H$\beta$ radial
surface emissivity distribution $F(r)$, and thereby the local
responsivity $\eta(r)$, and which for an assumed BLR geometry, result
in differences in the emission-line luminosity $L$, mean response
timescale $<\tau>$ and time-averaged responsivity $<\eta_{\rm eff}>$.
These include: differences in the range in (i) hydrogen gas density,
and (ii) hydrogen column density, (iii) allowing for extra-thermal
line widths, and (iv) differences in the ionizing continuum
normalisation. For expediency, we use photoionization grids previously
constructed with this wide range in parameters using Cloudy C90.04
(Ferland et al. 1998), again normalised to match the ionizing
continuum luminosity of NGC~5548.  While these grids have been
constructed using an earlier version of Cloudy than in Figures 1a, b,
here we are primarily interested in how changes in key model
assumptions affect the luminosity, delay and emission-line
responsivity, factors which are intimately connected to the
emission-line transfer function.\footnote{The SED used for the C90.04
  grids is described in Korista and Goad (2000, 2004), and is
  significantly harder than the Matthews and Ferland (1987) generic
  quasar continuum adopted for the C08 model grids, with a mean
  ionizing photon energy which is a factor of three larger.  It is for
  this reason that the diagonal ``cliff'' in EW, representing the $\log
  N(H) = 23$ clouds becoming fully ionized, is shifted by about $-0.5$
  dex in $\log Uc$ in Figure~3a (upper panel) compared to Figure~1a.
  However, we note that the Balmer emission lines are not particularly
  sensitive to the details of the ionizing continuum shape.}

We show in Figure~3a the EW(H$\beta$) (relative to the incident
continuum flux at 1215\AA) as a function of $n_{\rm H}$, $\Phi_{\rm
  H}$ for three choices of cloud hydrogen column density $N_{\rm H}$,
as well as for the default cloud column density $\log N_{\rm
  H}$~(cm$^{-2}$)$=23$ with the addition of a significant
micro-turbulent velocity for all clouds in the grid.  Contours and
symbols are as for Figure~1a. In each panel we show the nominal
boundaries in parameters contributing to the emissivity functions
$F(r)$ that appear in Figure~3b (see discussion in \S2.1).

\subsubsection{Gas hydrogen number density $n_{\rm H}$}

The radial surface emissivity distribution $F(r)$ is found to be
marginally sensitive to our choice of upper bounds for the
distribution function in the hydrogen gas number density, $n_{\rm H}$.
Figure~3b, panel (i), illustrates the effect of extending the upper
bound in $n_{\rm H}$ ($n_{\rm H}$(max)) from $10^{12}$~cm$^{-3}$
(solid black line) to $10^{13}$~cm$^{-3}$ (dashed red line).  For a
fixed value in $\log \Phi_{\rm H}$, increasing $n_{\rm H}$ increases
the H$\beta$ emissivity due to increased contributions from
collisional excitation.  Increased $n_{\rm H}$ results in enhanced
emission over the full radial extent of the fiducial BLR
(Figures~3b(i) left panel) and is particularly effective at enhancing
the emission measure at smaller BLR radii. When integrated over our
fiducial BLR geometry, the net result is a $\approx 20$\% increase in
the H$\beta$ luminosity, a marginally smaller responsivity-weighted
radius, and locally an enhanced emission-line responsivity
(Figure~3b(i), right-hand panel - dashed red line, and Table~1). The
enhanced responsivity arises due to the small increase in slope of the
radial surface emissivity distribution for BLR radii less than
$\approx$ 25 light-days that contains higher density gas with greater
efficiency in converting ionizing photons into Balmer emission line
photons.

\begin{figure}
\resizebox{\hsize}{!}{\includegraphics[angle=0,width=8cm]{fig3a.ps}}
          {{\bf Figure~3a.} As for Figure~1a, contours of $\log$(EW)
            for broad H$\beta$ referenced to the incident ionizing
            continuum at $\lambda$1215\AA\, for full source coverage,
            as a function gas hydrogen density $n_{\rm H}$ and flux of
            hydrogen-ionizing photons $\Phi_{\rm H}$. The smallest
            contour corresponds to 0.1\AA\; each solid line represents
            1 decade, and dotted lines represent 0.2 dex
            intervals. The solid star marks the old "standard BLR"
            parameters, the solid triangle the peak EW.  Model grids
            have been computed here using Cloudy version
            C90.04. Individual panels show the effect on the line EW
            of changing the adopted column density $N_{\rm H}$ from
            $\log N_{\rm H}$~(cm$^{-2}$)$=23$ upper left, to $\log
            N_{\rm H}$~(cm$^{-2}$)$=22$, upper right, and $\log N_{\rm
              H}$~(cm$^{-2}$)$=24$ (lower left) (see text for
            details). The lower right panel indicates the effect of
            introducing extra-thermal line widths, here in the form of
            a micro-turbulent velocity component $v_{\rm
              turb}=100$~km~s$^{-1}$, for a fixed hydrogen column
            density $\log N_{\rm H}$~(cm$^{-2}$)$=23$.}
\label{pgplot_c9004}
\end{figure}

\addtocounter{figure}{-1}

\begin{figure}
\resizebox{\hsize}{!}{\includegraphics[angle=0,width=8cm]{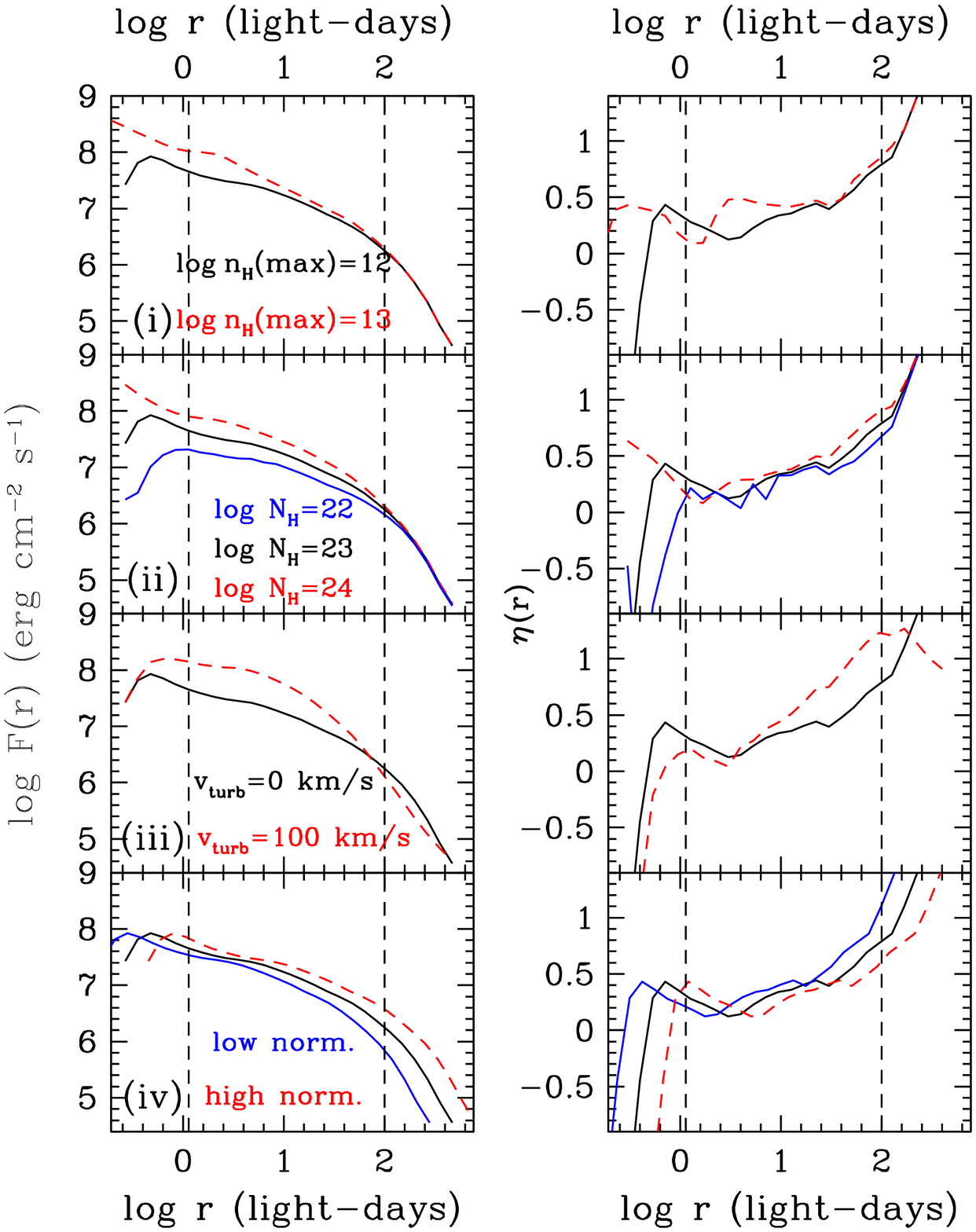}}
          {{\bf Figure~3b.} Left panels -- Radial surface emissivity
            distributions for broad H$\beta$ for an LOC model of
            NGC~5548. Individual panels show the effect of (i)
            changing the range in gas hydrogen densities $n_{\rm H}$
            from $8 < \log n_{\rm H}$ (cm$^{-3}$)~$ < 12$ (black line)
            to $8 < \log n_{\rm H}$ (cm$^{-3}$)~$ < 13$ (dashed red
            line), (ii) changing the cloud hydrogen column density
            $N_{\rm H}$ from $\log N_{\rm H}$~(cm$^{-2}$)$ = 23$
            (black line) to 22 (blue line) and 24 (red dashed line),
            (iii) increasing the micro-turbulent velocity $v_{\rm
              turb}$ from 0~km~s$^{-1}$ (black line) to
            100~km~s$^{-1}$ (dashed red-line), and (iv) changing the
            ionizing continuum normalisation by a factor 8.2 from a
            low (solid blue line) to high (dashed red line)
            (see text for details). Right-hand panels -- the
            corresponding radial responsivity distributions
            $\eta(r)$.}
\label{plot_fluxes_hb}
\end{figure}

\onecolumn
\begin{table}\label{tab1}
\centering
\caption{The link between emission-line luminosity, characteristic size and responsivity in the steady state.}
\begin{tabular}{@{}ccccc@{}}
\hline 
\multicolumn{5}{c}{Cloudy version C08} \\
H$\beta$           & $\log L$ & $R_{\rm LW}$ &  $R_{\rm RW}$ & $\eta(r=R_{RW})$ \\
                   & erg~s$^{-1}$ & (light-days) &  (light-days) &  \\ \hline
$\log N_{\rm H}$~(cm$^{-2}$)$=23$ & 41.497 & 36.5 & 41.0& 0.55 \\
 (the fiducial model)               &        &      &       &     \\ \hline
   $R_{\rm out}=50$~light-days        & 41.235 & 23.6 & 25.3 & 0.42 \\
$\log \Phi_{\rm H}=20$ at $R=7.5$~light-days          & 40.895 & 18.0 & 20.2 & 0.56 \\
low continuum normalisation                &  41.220      & 33.5     & 38.6     & 0.68    \\
high continuum normalisation                & 41.729       &  38.5    &  41.6     & 0.42    \\ 
                &        &      &       &     \\ \hline
\multicolumn{5}{c}{Cloudy version C90.04} \\ 
 H$\beta$           & $\log L$ &  $R_{\rm LW}$  & $R_{\rm RW}$ & $\eta(r=R_{RW})$ \\
                   & erg~s$^{-1}$ & (light-days) &  (light-days) &  \\ \hline
$\log N_{\rm H}$(max)~(cm$^{-2}$)$=23$      & 41.540  &  37.2  & 41.8   & 0.49 \\
$\log n_{\rm H}$(max)~(cm$^{-3}$)$=13$      & 41.623  &  35.6  & 39.8   & 0.49 \\
$v_{turb}=100$~km~s$^{-1}$            & 41.768  &  29.2  & 35.1   & 0.82\\  \hline
$\log N_{\rm H}$~(cm$^{-2}$)$=22$       & 41.374  &  39.0  & 43.0   & 0.42 \\ 
$\log N_{\rm H}$~(cm$^{-2}$)$=24$      & 41.686  &  35.0  & 40.0   & 0.60 \\ \hline
\end{tabular}
\end{table}

\noindent For our adopted luminosity distance $D_{\rm L} = 77.6$~Mpc
for NGC~5548, the mean narrow-line subtracted H($\beta$) luminosity,
corrected for Milky Way extinction ($A_{v} = 0.055$, Schlafly and
Finkbeiner 2011) over the duration of the 13~yr ground-based
monitoring campaign, $\log L$(H$\beta$) (erg~s$^{-1}$) =
41.756. Unless otherwise stated, the continuum has been normalised so
that $\log \Phi_{\rm H}$ (photons~cm$^{-2}$~s$^{-1}$) = 20 at a BLR
radius $R=15$~light-days.

\twocolumn

\subsubsection{Cloud hydrogen column density $N_{\rm H}$}

The BLR is likely comprised of clouds with a mix of hydrogen column
densities.  Here we consider the effect of a range in gas hydrogen
column density on the line EW as a function of $\Phi_{\rm H}$, $n_{\rm
  H}$, for photoionization model grids with fixed total hydrogen
column density $N_{\rm H}$.  The most obvious effect of increased
cloud column density (Figure~3a, panels~(i)--(iii)), is a general
increase in the EW(H$\beta$) for clouds of larger ionization
parameters as the cloud column density increases. The steep decline in
EW(H$\beta$) running approximately diagonally across the hydrogen
number density-ionizing photon flux plane occurs as the cloud becomes
fully ionized in hydrogen.

In Figure~3b, panel (ii), we illustrate radial surface emissivities
$F(r)$ for 3 hydrogen column densities $N_{\rm H}$ spanning the range
appropriate for BLR clouds $22.0 < \log N_{\rm H}$ (cm$^{-2}$)~$ <
24.0$.  With all else equal, the lower column density (solid blue
line) leads to smaller values for the radial surface emissivity,
reducing the luminosity by a factor $\approx $1.5, and increasing the
responsivity-weighted radius by $\approx 1$ light-day.  As can be seen
in comparing the upper two panels of Figure~3a, {\em all else being
  equal\/} there are relatively fewer lower column density clouds
emitting efficiently in H$\beta$ -- especially for larger values in $U
= \Phi_{\rm H}/n_{\rm H}c$ -- and so especially at smaller radii.
This results in a flatter radial surface emissivity distribution for
clouds of lower column density and consequently smaller responsivity,
and a larger responsivity-weighted radius (Table 1).  The converse is
true for clouds with larger column densities (compare the upper and
lower left panels). For $\log N_{\rm H}$ (cm$^{-2}$)$=24$, the
luminosity is a factor $\approx$1.4 larger, the responsivity
$\approx 20$\% larger, and with a $\approx$~5\% drop in the responsivity
weighted radius (see Table~1).

\subsubsection{Extra-thermal line widths}

As pointed out in paper~{\sc i}, and in Korista and Goad (2004),
$\eta(r)$ can be significantly enhanced by allowing for extra-thermal
line widths within the BLR gas. These may be caused by, for example,
micro-turbulent velocities (Bottorff et al. 2002), velocity shears, or
significant electron scattering within the line emitting gas. The
resulting reduction in the central line optical depths increases the
line escape probabilities and results in enhanced emission,
particularly in lines from clouds that have large central optical
depth for local line widths dominated by thermal motions. For the
hydrogen recombination lines, these are typically clouds with higher
values in the incident ionizing photon flux, and so smaller distances
from the central ionizing source.  Note that the peak EW has moved
upward by $\approx$~1 dex in $\Phi_{\rm H}$.  Micro-turbulent
velocities tend to open out the EW contours on the $\Phi_{\rm H}$,
$n_{H}$ plane, so that EW contours which formerly were almost constant
with $\Phi_{\rm H}$, now tend to follow lines of nearly constant
values of $Uc$, a diagonal line in the $\Phi_{\rm H}$, $n_{\rm H}$
plane (e.g. notice the differences between the 10\AA\, contours shown
in Figure~3a, panels (i) and (iv)).

The effect on H$\beta$'s radial surface emissivity $F(r)$ of
increasing the micro-turbulent velocity within the gas cloud from
0~km~s$^{-1}$ (solid black line) to 100~km~s$^{-1}$ (dashed red line)
is shown in Figure~3b, panel~(iii).  As above, a significant
micro-turbulent velocity boosts the emission across the entire
fiducial BLR geometry, but particularly at smaller BLR radii where the
clouds are optically thick in the hydrogen lines.  Additionally, when
extra-thermal line widths are included, the line responsivity is
significantly larger for radii beyond approximately 10~light-days out
to the outer boundary.  For the fiducial BLR geometry,
micro-turbulence results in a significant (factor $\approx$1.7) increase in
the emission-line luminosity, reduces the responsivity-weighted radius
by $\approx 15$\%, and increases the line responsivity by
$\approx$~65\% (see Table~1). We note that while this effect results
in a better match to the observed L(H$\beta$), the reduction in
$R_{\rm RW}$ is smaller than required by the data, and the
time-averaged responsivity $<\eta_{\rm eff}>$ is too large.

\subsection{Continuum normalisation}


The line emissivity and responsivity distributions within the BLR are
sensitive to small changes in the adopted continuum
normalisation. Here we distinguish between two types of continuum
normalisation.  The first relates to uncertainties in the incident
ionizing photon flux at a specified radial distance for a BLR of fixed
spatial extent. The ionizing photon flux is normally estimated by
assuming a continuum SED which is then scaled by the observed
continuum flux at a measurable wavelength, for example 1350\AA\ or
5100\AA\ (rest frame), for an assumed luminosity distance and after
applying corrections for extinction within the Milky Way galaxy and
potentially also the host galaxy. Since $\Delta \log r = -0.5 \times
\Delta \log \Phi_{\rm H}$ for a given ionizing luminosity and SED,
small uncertainties in the continuum normalisation in this case will
shift $F(r)$ towards the left (for lower continuum normalisations) or
towards the right (for higher continuum normalisations), within the
confines of the BLR (see e.g. Figure~3b panel~(iv), solid blue and
red-dashed lines, respectively).  These differences will lead to
changes in line $L$, $R_{RW}$ and $\eta(r=RW)$ (Korista and Goad 2004;
Goad and Korista 2014).  To illustrate this point low and high
continuum normalisations corresponding to a factor $\approx 8$ range
in ionizing continuum flux (e.g. Figure 3b, lower left panel),
increases $L$(H$\beta$) by a factor $\approx$ 3, increases $R_{\rm
  RW}$ from 38.6--41.6 light-days, and decreases $\eta(r=RW)$ from
0.68--0.42 (see Table~1).  Figure~\ref{plot_hb_small_norm}~panel~(ii)
illustrates model broad H$\beta$ light-curves for low (solid green
line) and high (solid black line) continuum normalisations
corresponding to the same factor of $\sim$8 range in the ionizing
continuum normalisation (see Table 2 for details). In all cases the
inner and outer boundaries, $R_{\rm in}$ and $R_{\rm out}$, have been
kept fixed at their starting values.  For a BLR with static boundaries
a lower continuum normalisation appears to be favoured.


The continuum normalisation is also determined by the way in which the
incident ionizing photon flux $\Phi_{\rm H}$ maps to BLR radial
distance $r$. An example of this, is a particular choice of source
luminosity distance $D_{\rm L}$. Uncertainties in $D_{\rm L}$ will
alter the radial surface emissivity distribution and the radial scale,
but in a self-similar way, i.e., $R_{\rm out}/R_{\rm in}$ remains
invariant. This is equivalent to a reassignment of the radial scale in
Figure~1b. Thus while the continuum and emission line luminosities
along with the BLR size (lag) will change, the radial dependence of
the surface emissivity and responsivity distributions within the
confines of the BLR will not.  For lower continuum normalisations of
this second type, the BLR will be smaller, with a corresponding drop
in the emission-line luminosity (though the emission line EW remains
the same). We note here that though the BLR is now more compact,
$R_{\rm out}$ remains equal to $R_{\rm subl}$.  However, the emission
line response amplitude will be unchanged if the effects of geometric
dilution are weak.

In Figure~\ref{plot_hb_small_norm} panel (iii) we illustrate this
effect by reducing the mapping of $r$ onto $\Phi_{\rm H}$ by a factor
2, resulting in a factor 4 drop in the mean emission-line luminosity
(see Table 1)\footnote{While a large uncertainty in the mapping of the
  cloud--source distance $r$ to the incident hydrogen ionizing photon
  flux $\Phi_{\rm H}$ due to an uncertainty in the luminosity distance
  is unlikely, an effect of this nature may also arise if the ionizing
  continuum is highly anisotropic (e.g., Netzer 1987; Nemmen and
  Brotherton 2010) so that the ionizing continuum flux incident on BLR
  clouds, and that which we infer from the observed continuum flux are
  not the same. If BLR clouds are located at large polar angles, and
  we observe them from much smaller ones, then the measured emission
  line EWs will be artificially reduced if continuum anisotropy is
  important.}. Since differences in the chosen value of $D_{\rm L}$
alter the luminosities of both line and continuum alike, the
discrepancy between the measured and model emission line luminosities
(factor $\approx 2$ for H$\beta$) will remain the same.  The net
effect is thus a decrease in the responsivity-weighted radius ($17.7 <
R_{\rm RW}$~(light-days)~$< 21.1$), while the predicted time-averaged
emission-line responsivity ($<\eta_{\rm eff}>=0.53$) is similar to the
fiducial model (Table~1). Indeed setting aside the discrepancy in the
emission line luminosity, Figure~\ref{plot_hb_small_norm}, panel (iii)
indicates that a lower continuum normalisation of this type provides a
far better representation of the data in terms of response amplitude
and delay.

The fidelity of the reproduction of the majority of the observed
features in the H$\beta$ emission-line light curve is testament to the
validity of using the scaled optical continuum as a proxy for the
driving ionizing continuum.

\addtocounter{figure}{2}

\onecolumn
\begin{figure}
\resizebox{\hsize}{!}{\includegraphics[angle=270,width=8cm]{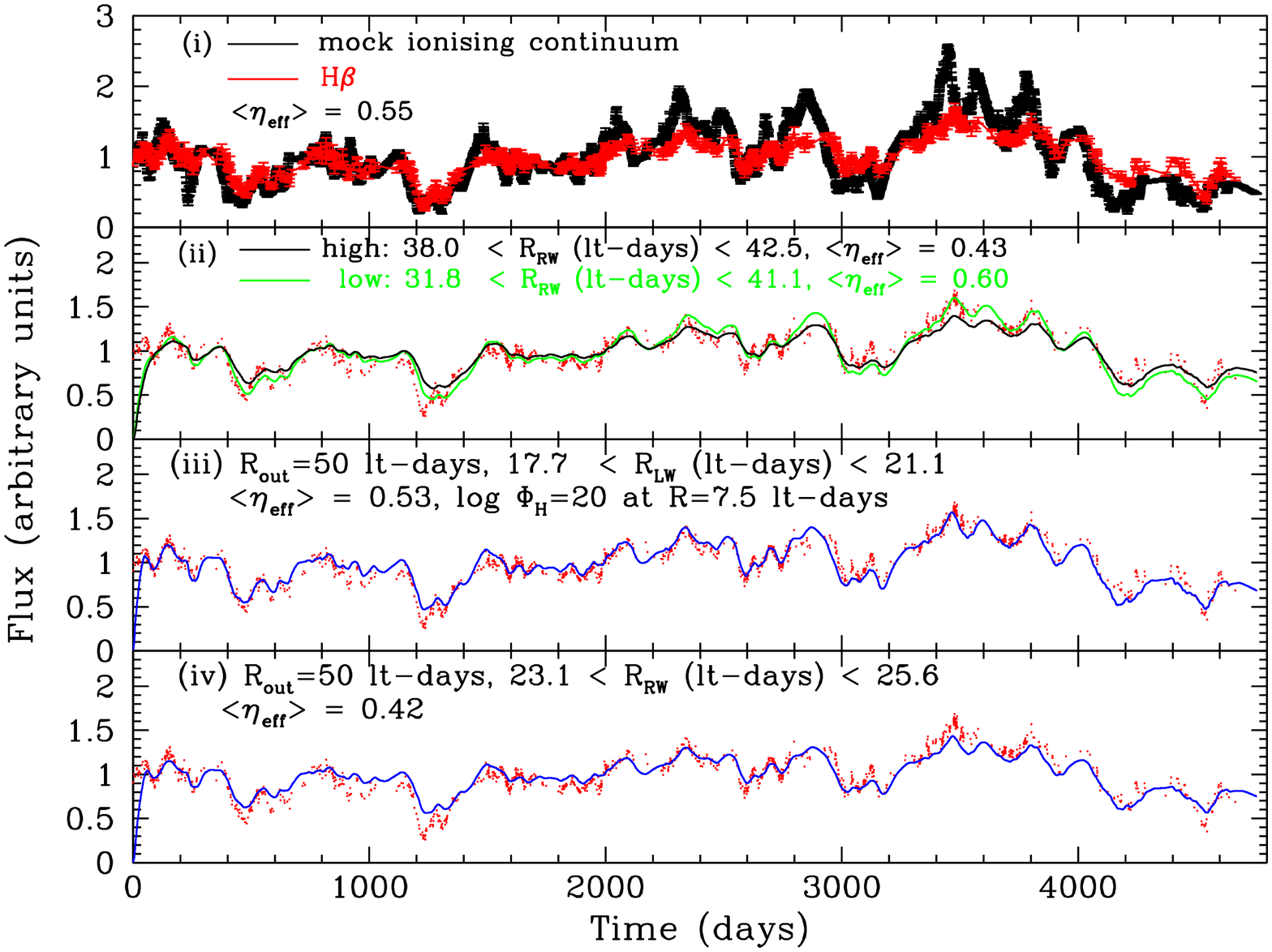}}
\caption{Model broad H$\beta$ emission-line light curves for our
  fiducial BLR geometry for (ii) high and low ionizing continuum
  normalisations (solid black and green lines respectively), (iii) a
  model BLR with a factor 2 reduction in the luminosity distance, and
  (iv) the fiducial BLR truncated at an outer radius of 50 light-days
  (solid blue line).  Both (iii) and (iv) result in a smaller BLR and
  lower line luminosities, but only (iii) results in enhanced
  amplitude response in the line. The quoted range in $R_{\rm RW}$
  corresponds to the range in measured CCF centroid for the 13 seasons
  of data.}
\label{plot_hb_small_norm}
\end{figure}

\twocolumn

\subsection{Is $R_{\rm out} \ll R_{\rm subl}$?}

We have also considered the possibility that our model BLR is simply
too large for our particular continuum normalisation.  The mean
response timescale for broad H$\beta$ for the fiducial BLR model,
determined from the centroid of the 1-d responsivity-weighted response
function $R_{\rm RW}$, is $\approx 41$ light-days. This is $\approx
20\%$ larger than the average continuum--emission-line delay of $35
\pm 1.7$~days determined for broad H$\beta$ (using our mock driving
ionizing continuum) over the 13~yr ground-based monitoring campaign of
NGC~5548.  While a smaller outer boundary will act to reduce the
responsivity-weighted radius of the BLR, and consequently reduce the
possible effect of geometric dilution (Goad and Korista 2014), it has
a number of undesirable consequences. First and foremost, the
emission-line luminosity is reduced. For the fiducial bowl-shaped BLR,
a factor 2 reduction in the BLR outer radius equates to a factor
$\approx$2 reduction in the emission-line luminosity for broad
H$\beta$. The luminosities of other emission-lines will also be
reduced though by differing amounts. Secondly, photoionization model
calculations suggest that for most emission lines, gas at larger radii
(or equivalently lower ionizing photon fluxes) will have the largest
responsivity (see Figure~1, lower panel, for H$\beta$). Thus the
emission-line responsivity in the absence of geometric dilution, when
averaged over the BLR cloud distribution, will tend to decrease if we
remove those contributions to the line responsivity arising from gas
at large BLR radii.  This is especially important for the geometry
considered here because as we have already shown (\S3.2), there is
little geometric dilution. Thus for the fiducial BLR geometry, the
variability amplitude of a particular line in response to continuum
variations is determined almost exclusively by the emission-line
responsivity for that line. {\em Note that if the local responsivity
  is small, the amplitude of the variations would be similarly small
  (for the same driving continuum), regardless of how compact the BLR
  is\/}.

Figure~\ref{plot_hb_small_norm} panel (iv) illustrates the effect of
truncating the outer BLR at a radius of 50 light-days.  Removing the
high responsivity gas at large BLR radii reduces the emission line
luminosity, the emission-line delay and its time-averaged
responsivity.  Thus when compared to the fiducial model (Figure~2
panel (iv)), the broad emission-line is less luminous (by a factor
$\approx$ 2), arises from a more compact region ($23.1 < R_{\rm
  RW}$~(light-days)~$< 25.6$), and has a smaller amplitude response in
the line ($<\eta_{\rm eff}>=0.42$ c.f. 0.51).  Once again, these
results indicate the deep connection between $L$, $R_{RW}$ and $\eta$.

We next compare this truncated BLR with one of the same size and outer
radius, but different radial surface emissivity distributions $F(r)$
arising from different mappings of $r$ onto $\Phi_{\rm H}$.  When
compared with Figure~\ref{plot_hb_small_norm} panel (iii), the
truncated BLR has a lower amplitude response in the line, because for
this line, the more responsive gas which is normally found at lower
ionizing photon fluxes (i.e. larger BLR radii) has now been removed.

That the emission-line luminosity, characteristic size and responsivity are 
intimately connected and cannot be treated in isolation is again
evident. Altering the models to address just one of these can
adversely effect the others. These connections should provide strong
constraints on BLR models, as well as help break the degeneracies
inherent in the interpretation of emission line transfer
functions. Consideration of additional emission lines will further
strengthen these tensions.

\section{The nature of the BLR inner and outer boundaries}

The models described thus far have been static, in the sense that the
spatial extent of the BLR has remained constant in time. These models
can still breathe, because in general the radial surface emissivity
distribution within the confines of the BLR inner and outer boundaries
is not a simple power-law, and consequently the local responsivity and
hence responsivity-weighted radius will vary with continuum level.
However, for the fiducial model the BLR is underluminous in H$\beta$
by a factor $\approx 2$, and is in general too large, while the {\em
  measured range\/} in responsivity appears too small, and in
particular fails to match the observed variations in H$\beta$ during
low continuum states.  Here, we investigate the possibility that the
BLR may in addition adjust its overall spatial extent in response to
changes in the ionizing continuum flux.

\subsection{The BLR inner boundary $R_{\rm in}$}

For the fiducial model the BLR inner boundary $R_{\rm in}$ has been
set to 200~$R_{\rm g}$ $\approx 1.14$~light-days for a $10^{8}$ solar
mass black hole. The location of this boundary was motivated in part
by the small measured delay in He~{\sc ii}~1640 in NGC~5548 (e.g.,
Korista et~al. 1995), although the precise location of the inner
boundary is unknown. However, we note that gas at such small radii has
very little surface area, and thus its contribution to the total power
of a particular emission line is modest at best. Additionally, at
small BLR radii the gas becomes over-ionized and the lines
thermalised.  Thus, unless the BLR is geometrically thin, then
provided that the specified inner boundary is small, relatively large
uncertainties in its location may be tolerated. In what follows unless
otherwise noted, we let the location of the BLR inner boundary vary
with continuum level according to $R_{\rm in} \propto C(t)^{1/2}$. For
H$\beta$ and other emission lines that form at large BLR radii, this
has almost no effect on the emission line variability.

\subsection{The BLR outer boundary $R_{\rm out}$}

For the fiducial model the location of the BLR outer boundary $R_{\rm
  out}$ is particularly significant because although the radial
surface emissivity distribution steepens as $r$ increases,
(Figure~1b), this fall in surface emissivity is (partially)
compensated for by a steady increase in the available surface
area. Here we have set $R_{\rm out}=100$ light-days, a distance beyond
which the mean continuum flux is low enough that robust dust grains
(e.g., graphite) can form and survive. When present ionizing continuum
and emission-line photons are destroyed on grains, and consequently
(along with the decrease in the gas phase abundances) the line
emission can drop significantly (Netzer and Laor 1993).

One consequence of a large surface area at large BLR radii is that
small changes in the location of the BLR outer boundary, whether
dynamical, or related to ionizing continuum variations, will produce
significant variation in the line emission at large radii.  This then
leads to significant variation in the emission-line delay and
amplitude of response, if gas exists at these radii. {\em Thus
  understanding the nature of the BLR outer boundary has become one of
  the key goals of AGN variability studies\/}.  In what follows we
investigate the behaviour of a dust-bounded BLR.

\subsection{A dust-bounded BLR model}\label{dusty}

Thus far our model BLR has been described by inner and outer
boundaries exterior to which the line emission is assumed to be zero
(i.e. a BLR which is truncated at both the inner and outer radius).
At small BLR radii this is likely to be a good approximation since:
(i) the line emission decreases rapidly at the highest incident
ionizing photon fluxes, and (ii) for most geometries, the inner BLR
contributes very little to the overall emission (an area effect).
However, for the BLR outer boundary this assumption is likely invalid,
since the reservoir of gas feeding the BLR likely originates in the
dusty torus at still larger radii. Alternatively, a truncated BLR, for
which the outer radius lies well within the location of the hot dust,
may arise if gas at larger radii is largely shielded from a direct
view of the ionizing continuum source.

Here, we assume that the BLR extends all the way to the inner edge of
the dusty torus, i.e., the distance to the hot dust. Where grains are
able to form and survive, they can compete efficiently for ionizing
photons and act to suppress, though not extinguish, the formation of
lines. Thus bounding this model BLR there exists a dusty region where
the line-reprocessing efficiency is significantly reduced. At
sufficiently large incident photon fluxes, grains charge up and heat
up and eventually sputter and sublimate. The distance at which grains
sublime depends on their chemical composition (e.g., graphites
vs.\ silicates) and size distribution (e.g. Barvainis 1987). Where
grains are sufficiently depleted line formation will efficiently cool
the gas. Thus the BLR outer boundary could move outwards with
increasing continuum source luminosity. Larger grains are more robust
than smaller grains. Thus condensation onto grains during lower
continuum flux states is most likely to occur onto larger grains.
During these low continuum flux states the effective line-emitting
boundary of the BLR could move inwards\footnote{In the fiducial model
  the BLR gas occupies the surface of an approximately bowl-shaped
  geometry (\S2). The dust we assume to follow the extension of this
  geometry out to larger radii, starting from a radial distance
  $R_{\rm subl}$, the distance at which robust grains can form and
  survive.  Note, here that the location of $R_{\rm subl}$ is confined
  to the bowl-surface, and is time-dependent, sliding along the bowl
  surface in response to variations in the ionizing continuum flux,
  i.e. $R_{\rm subl}=R_{\rm subl}(C(t))$.}.  In the context of a
dust-bounded BLR we here refer to an ``effective BLR outer boundary'',
$R_{\rm eff}$, which represents the point beyond which the line
emission is significantly suppressed.  The rate at which $R_{\rm eff}$
moves in response to continuum variations is governed by two important
timescales, the dust condensation and sublimation timescales,
$\tau_{\rm cond}$ and $\tau_{\rm subl}$ respectively.

Here we distinguish between the microscopic dust formation and
destruction timescales, which for individual grains is of order
minutes--days, and the macroscopic (or global) dust formation and
destruction timescales which is of order months--years for grains
largely shielded within or beyond the BLR gas, and relevant to the
discussion here (e.g. Kishimoto et al. 2013, 2011; H\"{o}nig and
Kishimoto 2011).  If the macroscopic dust sublimation timescale
$\tau_{\rm subl}$ is very much longer than the macroscopic dust
condensation (formation) timescale $\tau_{\rm cond}$, then the
location of the BLR outer boundary will be largely insensitive to high
continuum states, but will tend to move inwards on average during low
continuum states. Conversely, if the dust condensation timescale is
long compared to the dust sublimation timescale, the BLR outer
boundary will tend to move outwards during high continuum states and
remain there, unless there is a significant period in which the
ionizing continuum source remains in a low state.


  As viewed from the ionizing continuum source, at a given instant
  in time, emission line gas located at the same radial distance will
  be responding to the same ionizing continuum state. Gas located at
  larger radial distances will be responding to different (prior)
  continuum states. Thus there will be some regions in which the
  efficiency of line formation is increasing, while in others it is
  decreasing, depending on the prior continuum history, bracketed by
  regions in which the line emission efficiency is not affected by the
  presence of dust.  Furthermore, when viewed by an external distant observer,
  gas located at the same radial distance will appear to be responding
  to different ionising continuum states due to differences in the
  total light-travel time (reverberation).  Thus the observed location of
  $R_{\rm eff}$ at a particular instant in time is determined by both
  the local gas--grain physics and reverberation effects within the
  spatially extended BLR.  $R_{\rm eff}$ is therefore better described
  as a ``fuzzy'' or ``soft'' outer boundary.

\begin{figure}
\resizebox{\hsize}{!}{\includegraphics[angle=0,width=8cm]{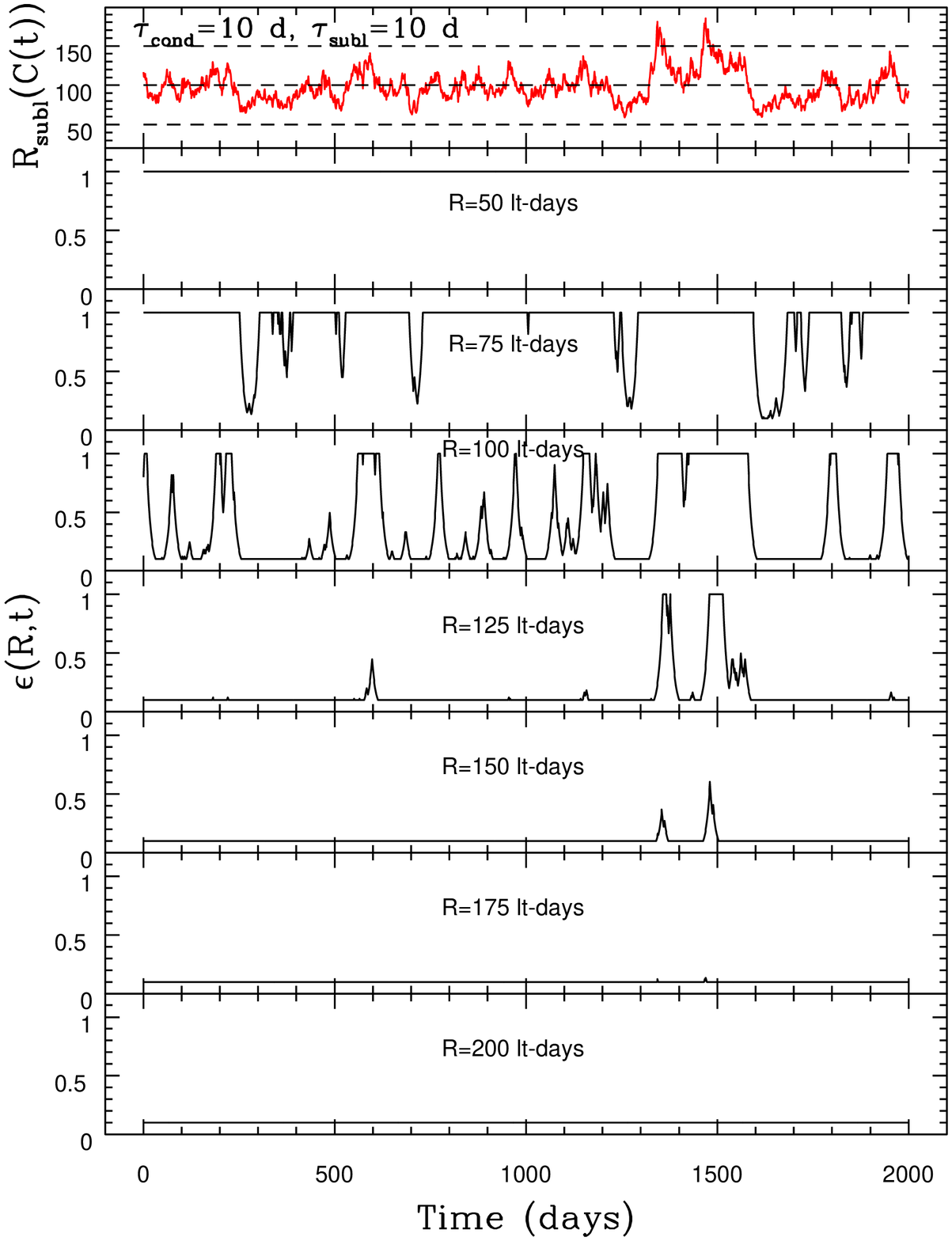}}
\caption{Upper panel - the predicted location of $R_{\rm subl}$
  as a function of time (days) for a dust-bounded BLR in which the
  location of the dust boundary varies as $C(t)^{1/2}$, and for which
  the local line re-processing efficiency factor $\epsilon(R,t)$
  depends upon the location of $R_{\rm subl}$, and the dust
  sublimation and dust condensation timescales, $\tau_{\rm subl}$ and
  $\tau_{\rm cond}$ respectively.  Shown are the results for
  $\tau_{\rm subl}=\tau_{\rm cond} = 10$~days. Panel 2--8, the
  re-processing efficiency factor $\epsilon(R,t)$ as a function of
  time (days) at fixed BLR radii $R$.  $\epsilon(R,t)$ displays rapid
  falls and sharp rises which appear symmetrical in shape.}
\label{dust_slice2}
\end{figure}

\begin{figure}
\resizebox{\hsize}{!}{\includegraphics[angle=0,width=8cm]{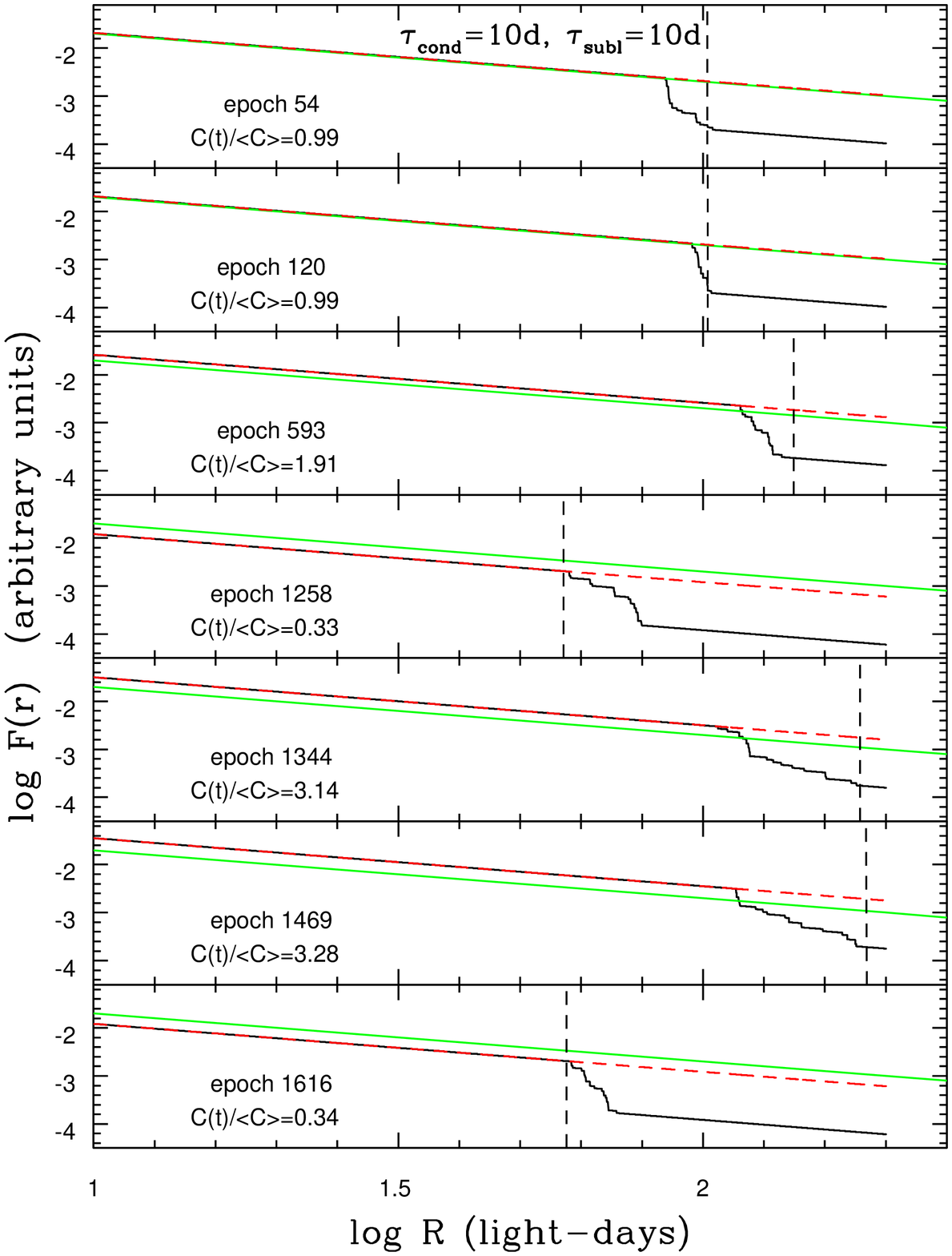}}
\caption{Snapshots of the instantaneous radial surface emissivity
  distribution $F(r)$ as a function of continuum level $C(t)/<C>$
  (solid black line), for the fiducial bowl-shaped BLR geometry
  described in section~\ref{dusty}. Also shown are the steady-state
  radial surface emissivity distribution (solid green line) together
  with the expected radial surface emissivity distribution assuming
  $\epsilon(R,t)=1.0$ (dashed red line).  The dashed vertical line
  represents the predicted location of $R_{\rm subl}$ for the epoch
  shown, assuming that the location of $R_{\rm subl}$ scales as
  $C(t)^{1/2}$. For this example, $\tau_{\rm subl} = \tau_{\rm
    subl}=10$~days. Since the dust sublimation and condensation
  timescales are equivalent, the location of the soft boundary $R_{\rm
    eff}$, as indicated by the sharp drop in the radial surface
  emissivity distribution (solid black line), is well-matched to the
  predicted location of the dust boundary for the current continuum
  level (dashed vertical line).}
\label{dust_emiss2}
\end{figure}

\subsubsection{A time-dependent efficiency factor $\epsilon(r,t)$}

In order to implement a dusty BLR outer boundary within the context of
our model, we introduce a time-dependent line-emission efficiency
factor $\epsilon(r,t)$ which we use to scale the radial surface
emissivity distribution $F(r)$, and which in the steady-state takes a
value of 1.0 in the absence of dust and a value of 0.1 when dust is
present. Initially we set the outer boundary $R_{\rm out}$ to be
located at the distance to the hot dust $R_{\rm out}=100$~light-days,
equivalent to the radius at which the temperature falls below the dust
sublimation temperature $R_{\rm out} = R_{\rm subl}$.  Thus in the
steady-state $\epsilon(r,t)= 1.0$ for $r \le R_{\rm subl}$, and
$\epsilon(r,t)= 0.1$ for $r > R_{\rm subl}$.  As the continuum varies,
the location of the dust boundary is assumed to scale as $R_{\rm subl}
\propto C(t)^{1/2}$ (as indicated by the solid red line in the
  upper panel of Figure 5, 7), similar to the relationship between
the distance to the hot dust and source luminosity found among a
sample of nearby AGN (Suganuma et al. 2004; 2006)\footnote{The
  relationship between the continuum luminosity and the inferred
  distance to the hot dust is found to be far shallower in individual
  sources, possibly as a consequence of long dust condensation and
  dust sublimation timescales when compared to the characteristic
  continuum variability timescale $T_{\rm char}$.}. NB the form
  of the driving continuum light-curve is similar to that illustrated
  by the solid red line in the upper panel of Figures~5, 7 and may be
  reconstructed by normalising this curve to its mean value and
  squaring the amplitude.

Here, we assume that as the continuum flux rises the line
re-processing efficiency factor $\epsilon(r,t)$ for gas lying interior
to the current location of $R_{\rm subl}$, increases exponentially
from its current value until it reaches a maximum value 1.0 on a
timescale $\tau_{\rm subl}$, where $\tau_{\rm subl}$, the macroscopic
dust sublimation timescale, represents the timescale over which grains
(embedded within the BLR clouds) are significantly depleted by UV
photons. Thus, for a given increase in the continuum flux, clouds
which lie interior to the region bounded by $R_{\rm subl}$, will
either emit at 100\% efficiency, or their efficiency will grow
exponentially, according to $\epsilon(r,t)={\rm
  min}[\epsilon(r,t-1)e^{1/\tau_{\rm subl}},1.0]$.  Conversely, as the
continuum flux decreases, we assume that for gas formerly bounded by
$R_{\rm subl}$ but now lying exterior to $R_{\rm subl}$, the line
re-processing efficiency factor $\epsilon(r,t)$ decreases
exponentially from its current value to a minimum efficiency factor of
0.1, on a characteristic timescale, $\tau_{\rm cond}$, the macroscopic
dust condensation timescale, such that $\epsilon(r,t)= {\rm
  max}[\epsilon(r,t-1)e^{-1/\tau_{\rm cond}},0.1]$.  Thus at any
instant in time, there will be a dust-free zone in which the line
re-processing efficiency $\epsilon(r,t)=1.0, \; \forall t$, a dusty
zone, in which $\epsilon(r,t)=0.1, \; \forall t$, bounding an
intermediate zone where grains are in the process of either being
formed or destroyed, and for which the line re-processing efficiency
lies in the range $0.1 < \epsilon(r,t) < 1.0$, and is either
decreasing exponentially to a minimum of 0.1 on a timescale $\tau_{\rm
  cond}$ due to a decrease in the continuum flux or increasing
exponentially to a maximum of 1.0 on a timescale of $\tau_{\rm subl}$
for a rising continuum flux. In the limit of $\tau_{\rm subl}$,
$\tau_{\rm cond}$ very small, the location of the effective BLR outer
boundary $R_{\rm eff}$ varies in lock-step with the ionizing continuum
(i.e. $R_{\rm eff} \propto C(t)^{1/2}$). This we here refer to as a
variable hard boundary model.  Conversely, for $\tau_{\rm subl}$,
$\tau_{\rm cond}$ very large, the effective BLR outer boundary remains
essentially static and behaves in a similar fashion to the fixed
boundary LOC model explored in \S3.  The light-curve generated by this
model will resemble that of Figure~2 panel (iv), except that the model
is now computed beyond 100 light-days out to a maximum radius $R_{\rm
  max}$ (here set at 200 light-days) and the line re-processing
efficiency $\epsilon(r,t)$ changes abruptly either side of this
boundary (from 1.0 to 0.1). For $R_{\rm max}=200$ light-days the total
covering fraction is $\sim$~70\% for the fiducial bowl-shaped geometry
(c.f. 50\% at $R=100$ light-days, see Goad, Korista and Ruff 2012, for
details).

\subsubsection{A simple toy model}

To illustrate the general behaviour of a dust-bounded BLR and its
sensitivity to the dust sublimation and dust condensation timescales,
we have generated model emission-line light-curves using a driving
continuum light-curve which can be described by a damped random
walk\footnote{We note that a damped random walk has been found to
    be a poor match to the broad band variability behaviour of AGN
    observed by Kepler (Mushotzky et al. 2011), albeit in a small
    sample of objects and for light-curves which only probe timescales
    appropriate for the disc light-crossing time.}  in the logarithm
of the flux (e.g. Kelly et al. 2009; MacLeod et al. 2010; Goad and
Korista 2014, paper~{\sc i}) and the fiducial BLR model with a
power-law radial surface emissivity distribution with slope $-1$,
bounded by dust at its outer edge. That is, in the steady state
$R_{\rm subl}(C(t)=<C>=1)=100$ light-days. We have deliberately chosen
a powerlaw radial surface emissivity distribution, since in this case
$\eta(r) = constant\; \; \forall r$, in the steady-state, and thus
breathing can only occur via changes in the location of the BLR
boundaries.  The radial surface emissivity distribution is computed
out to a maximum radius $R_{\rm max}$ of 200 light-days, and in the
steady-state the line re-processing efficiency factor $\epsilon(R,t)$
is assumed to be 1.0 for gas lying interior to $R_{\rm subl}$, and 0.1
for gas lying exterior to $R_{\rm subl}$.

We consider two scenarios for a BLR with a dusty outer boundary. For
the first, we set the dust sublimation and dust condensation timescale
to be equivalent to one another, so that grains are depleted and
reform on the same timescale, here $\tau_{\rm
  subl}=\tau_{\rm cond}=10$ days.  For the second, we set the dust
sublimation timescale to be factor ten longer than the dust
condensation timescale, $\tau_{\rm subl}=100$ days, and $\tau_{\rm
  cond}=10$ days, so that while grains are depleted rather slowly,
they rapidly reform. For the latter, quoted values of $\tau_{\rm
  subl}$ and $\tau_{\rm cond}$ were chosen to suppress large amplitude
emission-line variations during high continuum states, while allowing
for a more compact BLR with a smaller responsivity-weighted radius,
during prolonged low continuum states.

\begin{figure}
\resizebox{\hsize}{!}{\includegraphics[angle=0,width=8cm]{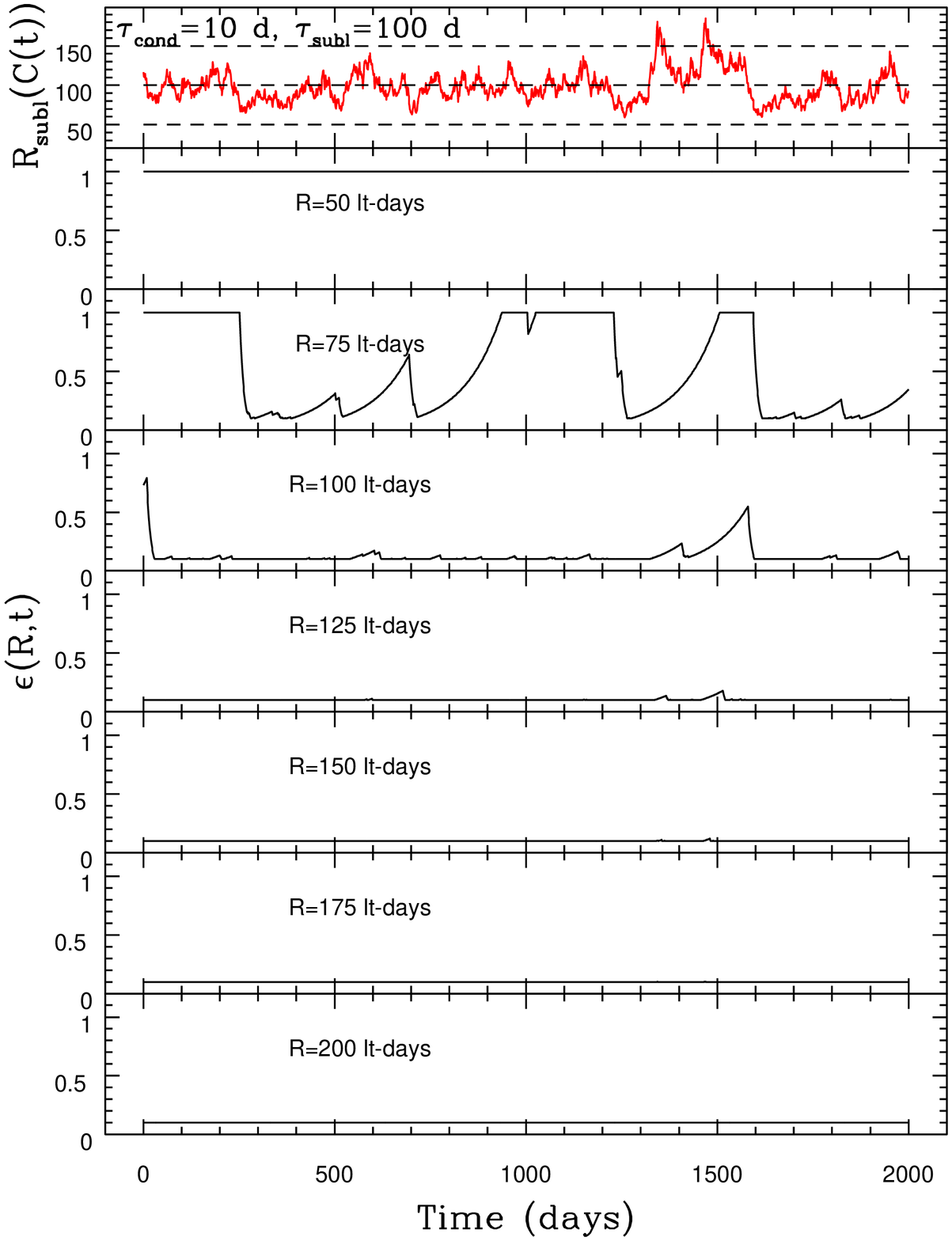}}
\caption{Upper panel - the predicted location of $R_{\rm subl}$
  (solid red line) as a function of time (days) for a dust-bounded BLR
  with $\tau_{\rm cond} = 10$~days and $\tau_{\rm subl}=100$~days.
  Panels 2--8, the model re-processing efficiency factor
  $\epsilon(r,t)$ as a function of time (days), for BLR radii of 50,
  75, 100, 125, 150, 175 and 200 light-days respectively. For this
  model, the time-dependent line re-processing efficiency factor shows
  rapid declines followed by slow rises.}
\label{dust_slice1}
\end{figure}

\begin{figure}
\resizebox{\hsize}{!}{\includegraphics[angle=0,width=8cm]{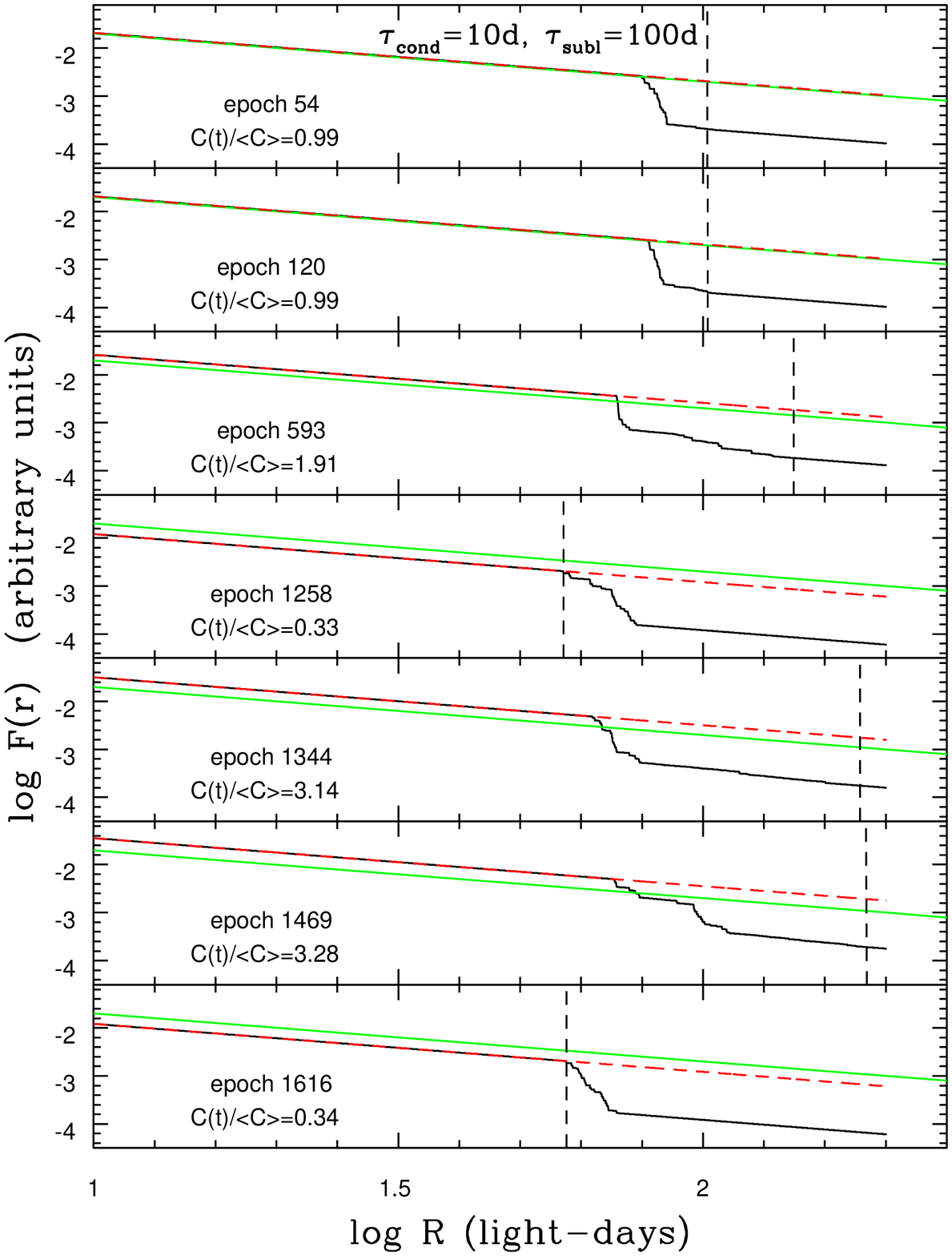}}
\caption{As for Figure~\ref{dust_emiss2}, adopting $\tau_{\rm cond} =
  10$~days, and $\tau_{\rm subl}=100$~days.  A strong hysteresis is
  evident in the radial surface emissivity distributions such that the
  location of the soft boundary $R_{\rm eff}$ (as indicated by the
  sharp drop in radial surface emissivity distribution) correlates
  only poorly with the predicted location of $R_{\rm subl}$ (as
  indicated by the vertical dashed line) for this continuum level.}
\label{dust_emiss1}
\end{figure}

The results of our simulations are shown in
Figures~\ref{dust_slice2}--\ref{dust_emiss1}. Each model is bounded at
large radii by $R_{\rm subl}$ which as for the variable hard boundary
model scales as $C(t)^{1/2}$ (Figure~\ref{dust_slice2}, upper
panel). However, since the line re-processing efficiency is low for
radii $R_{\rm subl} < r < R_{\rm max}$, the effective outer boundary
$R_{\rm eff}$, as indicated by the sharp drop in line emissivity
(figure~\ref{dust_emiss2}, solid black line) at larger radii, is
smaller than $R_{\rm max}$. Note that the range in radii over which
changes in the re-processing efficiency occur is extensive, spanning
$\approx$60-185 light-days, appropriate for the $\sim$ factor of 9.6
range (max/min) in continuum level.


For the first simulation, we set $\tau_{\rm subl}=\tau_{\rm
  cond}=10$~days.  Consequently, the temporal behaviour of the
re-processing efficiency $\epsilon(r,t)$ at fixed radial position $r$
is characterised by symmetric rises and falls, and by significant
excursions in re-processing efficiency on relatively short timescales
(Figure~\ref{dust_slice2}, panels 2-8).  In Figure~\ref{dust_emiss2}
we show snapshots of the instantaneous radial surface emissivity
distribution $F(r,C(t))$ (solid black line) at seven epochs, chosen to
illustrate a broad range in continuum level.  Also shown is the
steady-state radial surface emissivity distribution (solid green line)
together with the radial surface emissivity distribution at the
current epoch, assuming $\epsilon(r,t)=1.0\; \; \forall r$ (dashed red
line).  One consequence of adopting similar dust sublimation and dust
condensation timescales is that the sharp drop in the radial surface
emissivity distribution $R_{\rm eff}$ more closely coincides with the
predicted location of $R_{\rm subl}$ for the concurrent value of the
continuum flux (Figure~\ref{dust_emiss2}, vertical dashed lines).

For the second simulation $\tau_{\rm subl}$ is a factor of 10 longer
than $\tau_{\rm cond}$. Variations in $\epsilon(r,t)$, are here
characterised by a rapid decline in the re-processing efficiency
during low continuum states followed by a more gradual increase in the
re-processing efficiency with increasing continuum level as the dust
is eroded (Figure~\ref{dust_slice1}).  The location of $R_{\rm eff}$
therefore decreases significantly on relatively short timescales
following a drop in continuum level, but moves outwards only very
slowly as the continuum level starts to rise.  Thus a strong
hysteresis in the location of $R_{\rm eff}$ is a defining
characteristic of models in which there is a strong mismatch between
the macroscopic dust sublimation and dust condensation timescales
(c.f. the location of the sharp drop in the solid black lines with the
dashed vertical lines in Figure~\ref{dust_emiss1}).

\section{A dust-bounded BLR model for NGC~5548}\label{ngc5548}

We now turn our attention to modelling the broad H$\beta$
emission-line light-curve in NGC~5548 assuming a dust bounded BLR with
time variable spatial extent.  For a BLR with either fixed or variable
boundaries, four case studies may be considered: (i) fixed $R_{in}$,
fixed $R_{out}$ (the default scenario), (ii) variable $R_{in}$, fixed
$R_{out}$, (iii) fixed $R_{in}$, variable $R_{out}$, and (iv) variable
$R_{in}$, variable $R_{out}$. However, as already mentioned, when the
surface area of the emitting region is taken into consideration, the
low emission measure arising from gas at small BLR radii (for this
geometry) suggests that a variable inner boundary has little impact on
the overall emission-line response. We have confirmed this supposition
via model simulations. Thus case (iii), variable $R_{out}$, and to a
lesser extent case (iv), variable $R_{in}$, $R_{out}$, are of
primary interest here.

  As before, we adopt the fiducial BLR model as our baseline model
  against which comparisons with dust-bounded BLR models will be
  made. The fiducial model is truncated at a fixed outer radius of 100
  light-days, a distance beyond which there is no contribution to the
  total line emission.  This we drive with our mock ionizing
continuum light curve for NGC~5548, generating a model emission-line
light-curve using the radial surface emissivity curve for H$\beta$
(Figure~1b) from Korista and Goad (2004), and assuming a full
non-linear response in the line.  The fixed boundary model, equivalent
to a dust bounded BLR for which the dust condensation and dust
sublimation timescales are infinitely long, and for which the
  contribution to the total line emission of dusty clouds is set to
  zero (Figure~9, upper panel), serves as a point of reference.  For
each continuum--model emission-line light-curve pair, we compute the
mean delay (CCF centroid and peak), and effective responsivity
$<\eta_{\rm eff}>$ over the full 13~yr light-curve, referenced to the
mock driving continuum. We also report the range in measured delays
and responsivity over time periods approximating the 13 seasons of
data for this source\footnote{Here we define a season as extending
  from the last data point of the previous season to the first point
  of the following season inclusive. This has negligible effect on the
  measured seasonal lags while ensuring that when extrapolating the
  light-curves between seasonal gaps, the light-curve is well
  behaved.}.  The former may be compared with the range in
responsivity-weighted radii measured from the centroid of the
instantaneous 1-d response functions. Results for all of the simulated
emission-line flux light curves presented in this work are summarised
in Table~2.

\subsection{A variable hard boundary model: $\tau_{\rm subl}$, $\tau_{\rm cond}$ small}

For illustrative purposes only, we first consider what we here refer
to as a variable hard boundary model for which the dust condensation
and dust sublimation timescales are assumed to be small ($\tau_{\rm
  cond}=\tau_{\rm subl}=1$~day), i.e. $R_{\rm out}$ is tied to $R_{\rm
  subl}$.  For this model BLR clouds simply switch on/off once the
continuum is of a sufficiently high/low level, a process which we
consider to be unphysical.  However, we include it here as it serves
to illustrate the most extreme range in variability (i.e. it shows the
largest range in $R_{\rm eff}$).  Gas interior to the current location
of $R_{\rm subl}$ will be fully emissive, while that beyond $R_{\rm
  subl}$ will emit at only 10\% efficiency, Here $R_{\rm subl}$ acts
as a sliding on-off switch for BLR clouds with the location of $R_{\rm
  subl}$ governed by the relation $R_{\rm subl} \propto C(t)^{1/2}$.

Comparing Figure 9, panel (ii) (a dust-bounded BLR) with Figure 9
panel (i) (the fiducial fixed boundary model), illustrates a number of
key attributes of a dust-bounded BLR.  First, the range in delays, as
measured from the seasonal data is considerably larger than for fixed
boundary models. Note that the predicted range in
responsivity-weighted radii is large $27.1 < R_{\rm
  RW}$~(light-days)$~< 55.5$, indicating a BLR which is a factor 2
smaller in low continuum states than in high continuum states. Second,
the time averaged effective responsivity is significantly larger
($\eta_{\rm eff}=0.81$ c.f. 0.51) when compared to fixed boundary
models, and exceeds that measured for H$\beta$ for the 13~yr
monitoring campaign ($<\eta_{\rm eff}> = 0.55$, see Table~2). The BLR is more
compact in low continuum states and responds more coherently and with
larger amplitude than does a fixed boundary model of similar
dimensions.  Consequently, the variable boundary model is a far better
match to low continuum states, for example epochs 400--600, and
1200--1400, than is a fixed boundary model.  The mean delay at the
start of the campaign is still too long however, possibly pointing to
a more compact BLR geometry than that used here, or alternatively, a
prolonged low continuum state, and thus a smaller $R_{\rm subl}$ prior
to the start of the 13~yr campaign (see end of \S6.2).  However, the
extreme variation in the location of the BLR outer boundary exhibited
by this model is less successful at reproducing the emission-line
response in the highest continuum states during the latter half of the
13~yr campaign (e.g. epochs 2000--4000). The larger surface area
available at larger BLR radii for this geometry coupled with the
relatively shallow radial surface emissivity distribution results in
enhanced variability over and above that which is observed for broad
H$\beta$ during high continuum states, and when compared to models
with fixed outer boundary.  A variable ``hard'' outer boundary with
$\tau_{\rm cond}$, $\tau_{\rm subl}$ small, thus appears prohibited by
the data.

\onecolumn
\begin{figure}
\resizebox{\hsize}{!}{\includegraphics[angle=270,width=8cm]{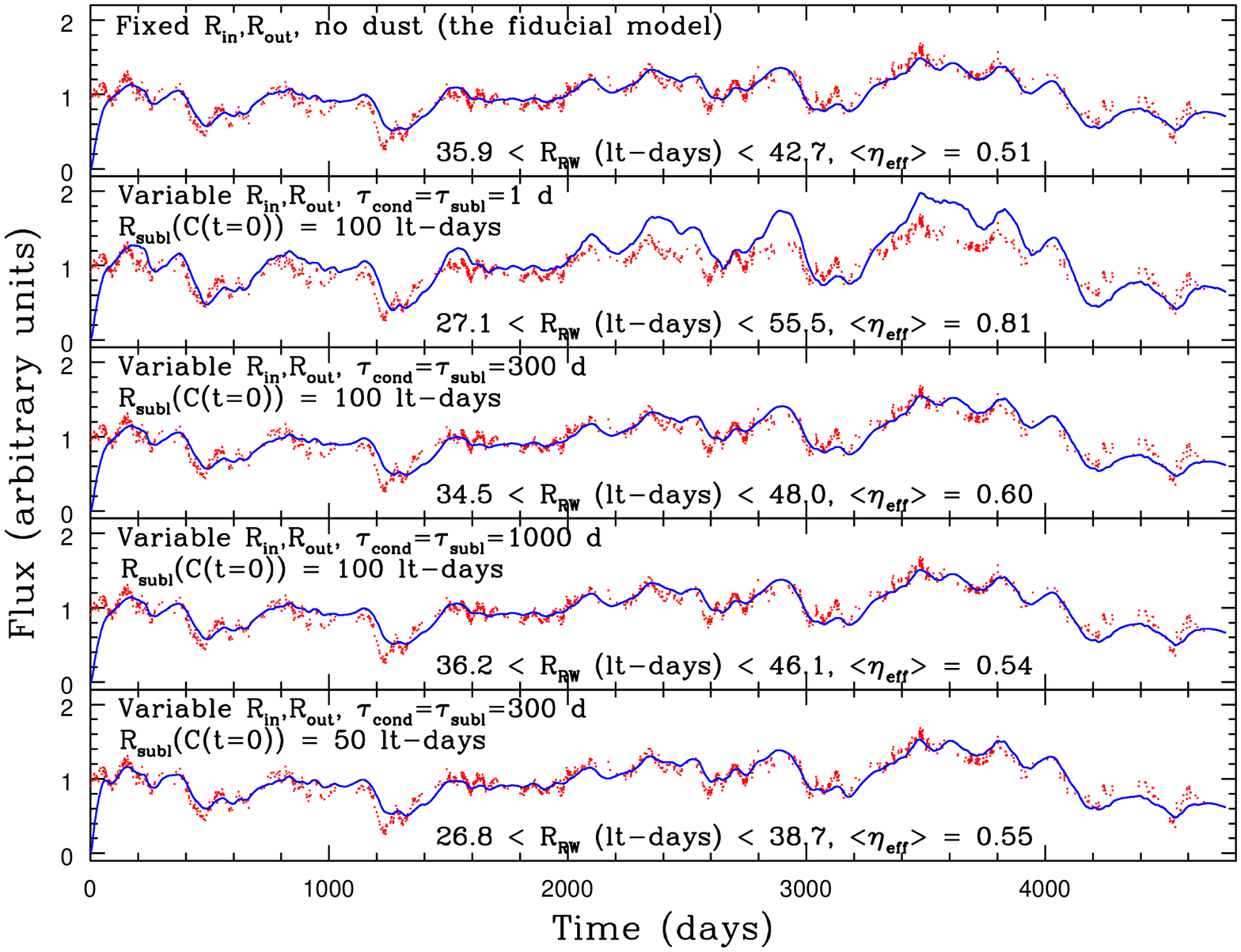}}
\caption{Model broad H$\beta$ emission-line light-curves for NGC~5548
  generated by driving the fiducial fixed boundary BLR model (upper
  panel) and variable dust bounded BLR models (panels 2--5) with the
  mock ionizing continuum. The observed H$\beta$ light-curve is
  indicated in red, simulated light-curves in blue. Quoted values are
  for the simulated light-curves.}
\label{hb_dusty}
\end{figure}

\twocolumn

\onecolumn
\begin{figure}
\resizebox{\hsize}{!}{\includegraphics[angle=270,width=8cm]{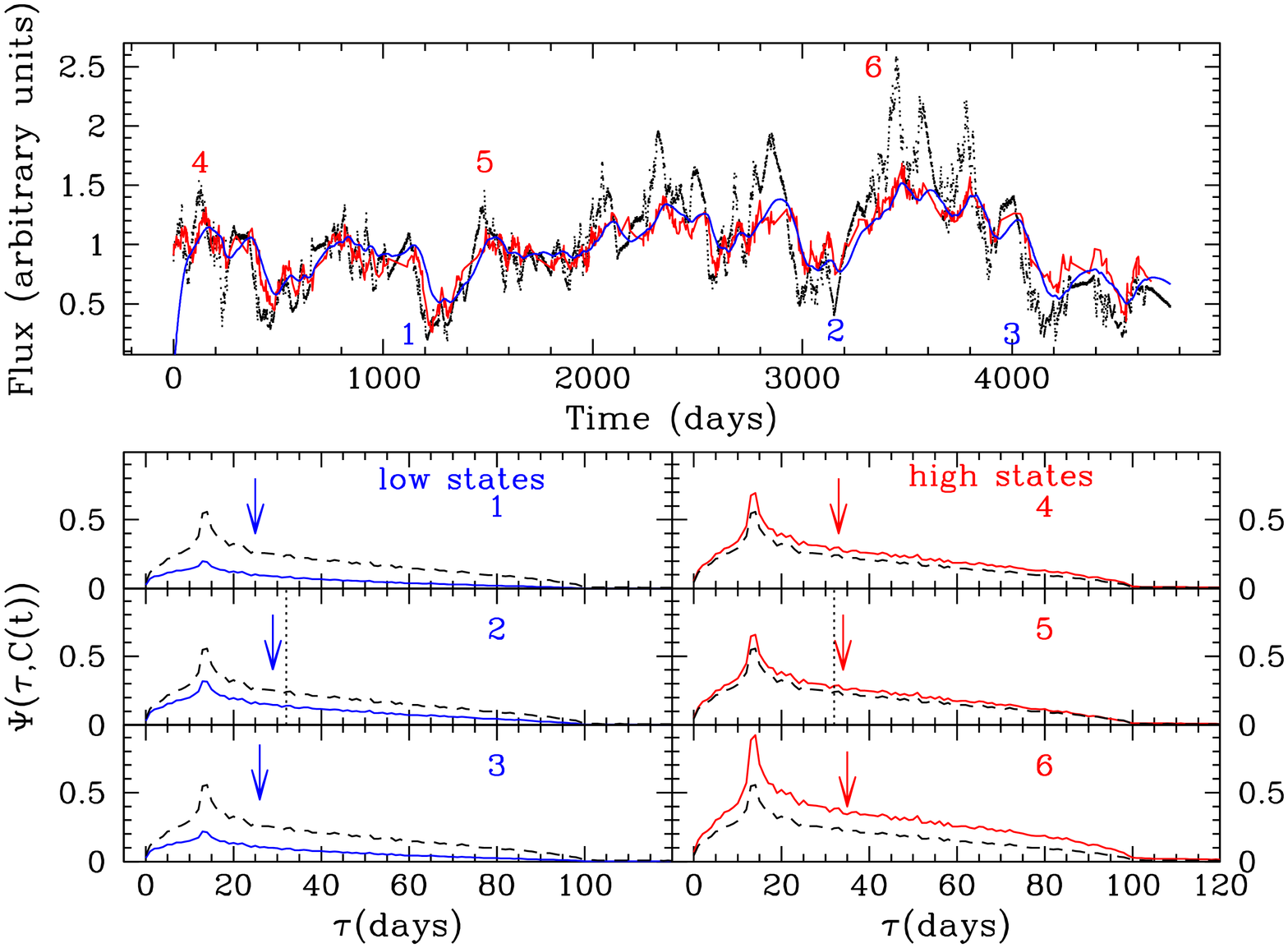}}
\caption{Upper panel - model continuum light curve (black dots), and
  observed (red line) and model broad H$\beta$ (blue line)
  emission-line light curves for a dust-bounded model of NGC~5548 with
  $\tau_{\rm cond}=\tau_{\rm subl}=1000$~days. Lower panels -
  corresponding instantaneous time-dependent transfer functions
  $\Psi(\tau,C(t))$, for 3 low-states (panels 2-4, LHS, solid blue
  lines) and 3 high-states (panels 5-7, RHS, solid red lines). For
  reference the steady--state transfer function corresponding to a
  mean continuum of 1 is indicated by the dashed black line. The
  vertical arrows indicate the centroid of $\Psi(\tau,C(t))$ at the
  corresponding epoch. For clarity, each has first been normalised to
  an arbitrary flux.  Note that the centroid of $\Psi(\tau,C(t))$ is
  large in high continuum states and small in low continuum
  states. The centroid of the steady state transfer function is
  indicated by the vertical dotted line in the middle 2 panels.}
\label{plot_response}
\end{figure}
\twocolumn

\subsection{A soft (or ``fuzzy'') BLR outer boundary}

For the fiducial BLR geometry, a variable hard boundary model, while providing
a better match to low continuum states, is found to be too responsive
during high continuum states, producing variability over and above
that which is observed.  We now consider dust bounded models in which
the dust condensation and dust sublimations timescales are comparable
to, or significantly longer than, the characteristic timescale of the
driving continuum light-curve. These we refer to as ``soft'' boundary
models.  Values of $\tau_{\rm cond}$, $\tau_{\rm subl}$ were chosen to
be large enough to suppress the excessive variability exhibited by
variable hard boundary models in high continuum states while still allowing
significant changes in $R_{\rm eff}$ during low continuum states.  We
note in passing that as $\tau_{\rm subl}$ and $\tau_{cond}$ increase,
the effective BLR size and its response become increasingly sensitive to the
long term history of the continuum variations.
First, we consider a model with a starting dust sublimation radius of
$R_{\rm subl}=100$ light days and with $\tau_{\rm
  subl}=\tau_{\rm cond}=300$~days (Figure~~\ref{hb_dusty}, panel
(iii)), a factor few larger than $\tau_{max}$ the maximum BLR delay at the
starting outer radius, and a close approximation to the characteristic
variability timescale of the mock driving continuum light-curve
(Kelley et al. 2009).  Figure~5 indicates that models for which
$\tau_{\rm subl} = \tau_{\rm cond}$ are characterised by symmetric
excursions in line reprocessing efficiency $\epsilon(r,t)$. Thus the
effective outer boundary will vary with continuum level with a delay
set by the dust sublimation and dust condensation
timescales. 

Figure~\ref{hb_dusty} panel (iii) shows that for $\tau_{\rm
  subl}=\tau_{\rm cond}=300$~days, the emission-line responsivity,
$<\eta_{\rm eff}>=0.60$, is somewhat larger than that measured over
the 13~yr campaign ($<\eta_{\rm eff}>=0.55$), but importantly,
significantly smaller than that measured for the variable hard
boundary model.  This general trend of decreasing emission-line
variability amplitude with increased $\tau_{\rm cond}$, $\tau_{\rm
  subl}$ has been verified with simulations (e.g. compare
Figure~~\ref{hb_dusty}, panels (ii)--(iv), and see Table~2). The
H$\beta$ emission-line variability amplitude for $\tau_{\rm
  subl}=\tau_{\rm cond}=300$~days, is generally larger than is
observed during high continuum states. This is a consequence of the
line emitting region extending to include high responsivity gas at
larger radii (see Fig~1b).  Furthermore, with such long timescales for
dust destruction and dust reformation, the range in spatial
extent of the BLR $34.5 < R_{\rm RW}$~(light-days)~$ < 48.0$, is found
to be only marginally larger than that found for a model with static
BLR boundaries ($35.9 < R_{\rm RW}$~(light-days)~$ < 42.7$). 

With such large values for $\tau_{\rm subl}$ and $\tau_{\rm cond}$,
the spatial extent of the BLR will remain large on average except
during prolonged periods (longer than the BLR light-crossing time) of
low continuum flux. In the absence of prolonged low continuum
states, a more compact BLR may be realised if $\tau_{\rm cond}$ is
significantly shorter than $\tau_{\rm subl}$. Under these circumstances,
the BLR will become more compact during the decline toward lower
continuum states but will not grow in size as quickly when
transitioning toward higher continuum states (e.g., note the
difference in the observed decline in $F(r)$ (i.e. $R_{\rm eff}$)
relative to the expected location of $R_{\rm subl}$ for the low (panel
  (iv)) and high (panel (vi)) continuum states shown in Figure~8). The
  BLR will thus be smaller on average than for a model in which
  $\tau_{\rm cond} \ge \tau_{\rm subl}$.

In Figure~~\ref{hb_dusty} panel (iv) we illustrate the effect of
increasing the dust condensation and dust sublimation timescales to
1000~days, which from dust reverberation mapping experiments, are
thought to be representative of the likely macroscopic dust
condensation and dust sublimation timescales in nearby AGN (Kishimoto
et~al. 2013; Schn\"{u}lle et al. 2013, 2015). This model achieves the goal of
suppressing excessive variability in high continuum states, and has a
similar time-averaged responsivity ($<\eta_{\rm eff}>=0.54$) to the
fixed boundary model, but at the expense of a smaller range in $R_{\rm
  RW}$ than models with smaller $\tau_{\rm cond}$, $\tau_{\rm subl}$. 
For this model, large changes in the effective outer boundary will
only become apparent for prolonged rises or falls in the ionizing
continuum flux\footnote{We note that long dust sublimation and dust
  condensation timescales will introduce a memory into the system
  behaviour (other than that attributed to reverberation effects
  within the spatially extended BLR) that may persist well beyond the
  characteristic timescale of the driving continuum light-curve or
  indeed the maximum delay at the BLR outer radius.}.

With the caveat that we are here exploring the behaviour of only one
of many possible BLR geometries, these simulations suggest that if the
BLR is bounded by dust at its outer edge, then dust condensation and
sublimation timescales must be comparable to, or significantly longer
than the characteristic timescale of the driving continuum light-curve
$T_{\rm char}$, so that $R_{\rm eff}$ remains approximately constant
(e.g. compare panels (i) and (iv) of Figure~\ref{hb_dusty}).  We note
that none of the dust-bounded simulations discussed so far are
particularly successful at matching the observed line behaviour during
the low continuum state at 1200 days.  Either the BLR is more compact
than that described here, or $\tau_{\rm cond}$ may be significantly
shorter than $\tau_{\rm subl}$).

In the bottom panel of Figure~{\ref{hb_dusty} we simulate the
  emergence from a prolonged low continuum state prior to the start of
  the campaign, setting the starting dust sublimation radius $R_{\rm
    subl}(C(t)=0))$ to 50~light-days, for a dust-bounded BLR for which
  the dust sublimation and dust condensation timescales are 300
  days. This model was motivated by the relatively poor match of
  previous models to the first very low continuum state (epochs
  1200-1400), i.e. at the start of the campaign the BLR responds too
  slowly (the delays are too large), and the amplitude of response is
  a poor match to the observed emission-line strength during low
  continuum states.  If prior to the onset of the 13 yr ground-based
  monitoring campaign, the BLR had been in an extended low-state, then
  it is at least plausible that the dust extended to far smaller BLR
  radii than that considered here. This model exhibits some promising
  characteristics. The variability timescale is in general smaller
  than that for a static boundary model $26.8 < R_{\rm
    RW}$~(light-days)~$ < 38.7$, while the variability amplitude
  remains high $<\eta_{\rm eff}>=0.55$, though marginally less than
  that shown by a model with a larger initial radius for the hot dust
  (Figure~9 panel (iii)). It is also a better match to the observed
  short timescale variability and to the observed variability in high
  continuum states than a fixed boundary model with $R_{\rm
    out}=50$~light-days (Figure~4, panel (iv)).  However for such a
  small BLR, the H$\beta$  luminosity remains an issue.

In Figure~10 we compare the instantaneous continuum level dependent
transfer functions, $\Psi(\tau, C(t))$, for a dust-bounded BLR, with
$\tau_{\rm cond}=\tau_{\rm subl}=1000$~days, as determined for a
selection of low- (solid blue curve) and high- (solid red curve)
continuum states over the 13 yr campaign.  The centroid of $\Psi(\tau,
C(t))$ is indicated by the coloured arrows. The dotted vertical line
indicates the centroid of the transfer function for the steady-state
(i.e. $\Psi(\tau,C(t)=1)$, dashed black curve). The amplitude and
centroid of the continuum-level--dependent transfer function are
larger in high continuum states and smaller in low-continuum states.
Note that even though the low continuum states (labelled 1--3), are
similar in flux, the transfer functions (shown in blue) display
significant differences. This is a consequence of differences in the
prior continuum history which can lead to differences in the effective
BLR size in the presence of dust for the same continuum level.

The general trend of decreased variability amplitude (particularly in
high-continuum states) with increasing $\tau_{\rm cond}$, $\tau_{\rm
  subl}$, suggests that if the BLR is bounded by dust, then the dust
sublimation and dust condensation timescales are long when compared
to the BLR ``size'' and characteristic timescale of the driving
continuum. Kishimoto et al. (2013) recently reported direct evidence
for a receding dust sublimation region, using near-IR interferometry
of the nearby type~{\sc i} AGN, NGC~4151. They find that the size of
the near-IR emitting region scales with the long-term average
UV/optical continuum flux, brightening with a delay relative to the
UV/optical continuum on timescales of a few years. This suggests that
the macroscopic dust sublimation timescale is at least of order a few
years in duration in this object in line with our
simulations. Similarly, in an independent study Schn\"{u}lle et
al. (2013) found that the radiation from the hot dust in NGC~4151
brightens as the continuum increases with a delay of $\sim
50$~days. They suggest that the hot dust is cooler than the
sublimation temperature and therefore lies beyond the dust sublimation
radius and is fairly robust to destruction by UV photons on short
timescales, again pointing toward a rather static dust distribution.

In summary, while allowing the location of the BLR outer boundary to
vary in response to continuum variations has some obvious advantages
(allowing a more compact BLR in low continuum states and a larger BLR
in high continuum states), these (advantages) are mitigated by the
excess variability produced during high continuum states. Our simple
toy model suggests that dust can suppress the variable contributions
to the line emission from the outer BLR, provided that the dust
sublimation timescales are long.  These simulations therefore appear
to favour a BLR outer boundary which is robust to significant
continuum variations (i.e. a static BLR outer boundary).

\subsection{What limits the BLR outer radius?}

The radial extent of the BLR is a key quantity because it defines the
volume within which the emission-line luminosity and variability must
ultimately be derived.  The BLR must be sufficiently large to
reprocess a substantial fraction of the source ionizing luminosity,
but cannot extend to arbitrarily large radii. The velocities at large
radii, and indeed the line-emissivities arising from such gas, would
be small enough that the emission-lines (if present) would be weak and
narrow.  For the fiducial geometry described here, the BLR outer
boundary has been set by the distance at which dust grains can form
(i.e. the location of the hot dust).  This choice was informed by
reverberation mapping of multiple broad emission-lines and dust within
individual sources, which suggest that the line emitting region
extends at least as far as the expected location of the hot dust
(Clavel et al. 1991; Krolik et al. 1991; Peterson et al. 2002, Barth
et al. 2011; Koshida et al. 2014). Moreover, since the BLR and
ultimately the accretion disc are likely supplied by the reservoir of
gas residing in the dusty torus, scenarios in which there exists a
substantial physical gap between the outer BLR and the inner edge of
the dusty torus seem physically less attractive.

While the observable line-emitting gas in the fiducial model is
approximated to lie along the surface of a bowl (which has a large
covering fraction for polar angles greater than 50 degrees), we do not
exclude the possibility that this surface may be ``patchy'' and that
gas located at significant depth behind the surface may also
contribute to the observed line emission. These shielded clouds will
be exposed to a heavily filtered ionizing continuum, and so may reside
in an environment suitable for significant dust formation.  The
fiducial model geometry is limited in extent along the upper reaches
of the bowl-surface by dust and, by construction, behind the surface
by severe cloud--cloud shadowing (and/or dust).  These effects may be
also important in limiting the BLR spatial extent for other very
different BLR geometries, or alternatively, the BLR could simply be
gas bounded.

\section{Discussion and Summary}\label{discuss}

In paper~{\sc i} we showed that for model BLRs of varying size and 
fixed boundaries, that the measured emission-line responsivity and
delay are correlated, for characteristic timescales of the driving
continuum light curve which are less than the maximum delay at the BLR
outer radius.  This we attributed to geometric dilution arising from
reverberation effects within the spatially-extended BLR which act to
reduce the emission-line responsivity and delay from their expected
values, but in a predictable way. Next, we showed that the measured
emission line responsivity and delay are sensitive to the duration of
the monitoring campaign, and less so on sampling rate (for a randomly
sampled light curve). Significantly, we also found that in order to
satisfy the observed {\em intrinsic\/} Baldwin Effect (Kinney, Rivolo,
and Koratkar 1990; Pogge and Peterson 1992; Gilbert and Peterson 2003;
Goad, Korista, and Knigge 2004; Kong et~al.\ 2006), and reproduce the
observed strong positive correlation between BLR size and luminosity
in a single source (Peterson et~al.\ 2003; Cackett et~al.\ 2006), the
line radial surface emissivity distribution $F(r)$ must steepen toward
larger BLR radii.

In the present work, we deliberately focused our attention on the
nature of the BLR outer boundary, $R_{\rm out}$, and in particular its
location in the presence of ionizing continuum flux variations, since
$R_{\rm out}$ sets the volume within which the total luminosity and
time variable nature (i.e., characteristic size and responsivity) of
the emission lines is determined.  This is especially true of emission
lines that form at larger BLR radii (e.g., H$\beta$ and Mg~{\sc ii}),
which are often used in scaling relationships applied to black hole
mass determinations at high redshift (McLure and Jarvis 2002;
Vestergaard and Peterson 2006; Bentz et al.\ 2009; Kilerci Eser et al. 2015).

However, the physical mechanism that sets the BLR outer boundary
remains uncertain.  It may be photon limited, dust-bounded or simply
truncated (i.e., one that cuts-off due to a particular geometrical
arrangement of gas).  It is often assumed that the location of $R_{\rm
  out}$ is set by the dust sublimation radius, heretofore with little
or no consideration for what might happen to this boundary when the
incident continuum varies.  In keeping with the idea that the BLR
extends from the outer accretion disc to the inner edge of the duty
torus (Netzer and Laor 1993; Nenkova et al.\ 2008; Mor and
Trakhtenbrot 2011; Mor and Netzer 2012; Goad, Korista and Ruff 2012),
we here have explored the effects of imposing a dust-bounded BLR on
$L$, $R_{\rm RW}$ and $\eta$ for a particular emission line,
H$\beta$. Observationally, the location of the hot dust has been shown
to vary with continuum luminosity, both within an individual AGN and
among the AGN population as a whole (Suganuma et al.\ 2004, 2006;
Koshida et al. 2014). Thus, in the present work, we have also
considered a time-variable location for the BLR outer boundary in
response to continuum variations. With this aim we constructed a
mock ionizing continuum light curve using the best available optical
continuum light curve from the 13 yr ground-based monitoring campaign
of NGC~5548, to drive emission line variations (see \S2.2).

With reference to the best-studied AGN, NGC~5548, if dust limits the
spatial extent of the BLR, significant correlations between the
continuum level and the effective outer boundary of the BLR are ruled
out, because the emission line lags become far too long and the gas
becomes overly responsive in the higher continuum states.
Dust-bounded BLR models therefore favour dust sublimation and
condensation timescales which are large compared to both the BLR light
crossing time and the characteristic variability timescale of the
driving continuum, also favoured in dust-reverberation experiments
(H\"{o}nig and Kishimoto 2011; Kishimoto et al.\ 2011, 2013;
Schn\"{u}lle et al.\ 2013, 2015).

A static BLR imposes strong constraints upon the physical properties
of the line emitting gas.  With an outer boundary set by the graphite
sublimation radius, the BLR model is underluminous in H$\beta$ by a
factor 2, and the delays are too large. Furthermore, although
the model response amplitude of H$\beta$ {\em is a good approximation
to the observed time-averaged value}, the model emission-line light
curve is a poor approximation to the observed flux variations during
the lowest continuum states.

The emission line delays may be reduced on average by choosing a
smaller cut-off for the BLR outer boundary (i.e., $R_{\rm out} <<
R_{\rm subl}$). However, this leads to a still lower predicted
luminosity, as well as a smaller amplitude response in H$\beta$
despite the BLR being more compact\footnote{A weak correlation between
  the continuum and the emission-line variations does not necessarily
  imply a large BLR. Rather, in the absence of significant geometric
  dilution it may be indicative of low reprocessing efficiency for
  that emission line (e.g. Mg~{\sc ii}). This may in part explain the
  difficulty in obtaining an accurate lag estimate for this line (see
  e.g., Clavel et al.\ 1991; Krolik et al.\ 1991; Cackett et
  al.\ 2015).}.  This smaller amplitude response arises because the
gas with largest responsivity, usually associated with low incident
ionizing continuum fluxes and equated with larger BLR radii, is no
longer present.  This is an illustration of the strong tensions
between $L$, $R_{\rm RW}$, $\eta$ for a particular emission line.
Since these three quantities are encapsulated in the emission line
transfer function, an understanding of the tensions between them may prove
useful in the interpretation of velocity-resolved delay signatures
recovered from reverberation mapping experiments. They may also be
used to inform the continued development of forward modelling
techniques, recently employed to map the geometry and kinematics of
the BLR gas and constrain black hole masses.

We also found that in general, a lower continuum normalisation
provides a better match to the emission-line delays and responsivities
(e.g., Figure 4, panels (ii) and (iii)), with the caveat that the
predicted H$\beta$ emission line luminosities for these models remain
too small (see Tables 1 and 2 for details).  Since large uncertainties
in the luminosity distance to NGC~5548 are unlikely, these simulations
may point toward another mechanism for altering the mapping of
$\Phi_{\rm H}$ onto $r$, i.e., a smaller $\Phi_{\rm H}$ for a given
$r$ than that inferred from the observed UV continuum flux and typical
models of the continuum SED.  As an alternative, we investigated
various ways of boosting the H$\beta$ emissivity $F(r)$ and
responsivity $\eta(r)$ at smaller BLR radii, which resulted in a
reduction in the responsivity-weighted radius without requiring
changes in the BLR geometry (see \S4, Figure~3a,b and Table~1).


%


The physical size of the BLR is determined in large part by the energy
deposited into and reprocessed by the system. The luminosity-weighted
radius of a given emission line is dictated by the distribution of the
continuum reprocessing efficiency for that line and the distribution
of cloud solid angle intercepting the ionizing photons.  That the
energy deposited is important in constraining the line-emitting
geometry, is amply demonstrated in Horne, Korista and Goad (2003,
their Figures~5, 6).  They show that even for a single cloud model,
whose reverberation signature is described by a paraboloidal iso-delay
surface (for which the cloud--source distance is unconstrained), the
correct radial distance may be recovered if the emitted energy in the
emission line is properly accounted for, for a specified incident
ionizing continuum flux. The connection between energy and BLR size is
also revealed through the {\em remarkably tight observed relation\/}
between the measured characteristic time-delay $\tau$(H$\beta$)
(interpreted as $R_{\rm BLR}/c$) and the observed luminosity (Bentz et
al. 2013; Kilerci Eser et al. 2015). The physical size scale for the BLR is
also revealed through the mass of the central black hole $M_{\rm BH}$,
via the virial relation $v^{2}R_{\rm BLR} \propto M_{\rm BH}$,
although $R_{\rm BLR}$ is again inferred from the
continuum--emission-line delay information. Without additional
information (e.g., energy), $R_{\rm BLR}$ remains degenerate in delay.

Finally, that the majority of features observed in the 13~yr H$\beta$
emission-line light curve are captured in the simulations (see Figure
4, panel (iii)) validates use of the scaled optical (or equivalently
the UV) continuum as a proxy for the driving ionizing continuum.  It
also demonstrates the power of photoionization models for gaining an
understanding of the BLR.

\onecolumn
\begin{table}\label{tab2}
\centering
\caption{H$\beta$ emission-line delays (CCF peak, CCF centroid) and time-averaged responsivities for the full 13 yr light-curve (columns 2--4), and for the seasonal light-curves (columns 5--8), see text for details.}
\begin{tabular}{@{}lccccccc@{}}
\hline
             &  $CCF_{\rm peak}$  & $CCF_{\rm cent}^{\star}$ & $<\eta_{\rm eff}>^{\dagger}$ & $CCF_{\rm cent}$ (range)$^{\ddagger}$ & $R_{\rm LW}$ &  $R_{\rm RW}$ & $\eta_{\rm eff}$ (range) \\
           &  (days)   & (days) &   &  (days) & (light-days) & (light-days) &  \\
                        &             &      &         &     &  &  &  \\          
observations            &  $17.3\pm 1.4$ & $35.6\pm 1.7$ &  0.55  &  12.4 -- 27.0  & NA           & NA           &     0.3 -- 0.99  \\
                        &             &      &         &     &  &  &  \\          
                        &             &      &         &     &  &  &  \\ \hline         
\multicolumn{8}{c}{$R_{\rm out} = $100~light-days, fixed boundaries} \\ \hline
                        &             &      &         &     &  &  &  \\          
power-law, $\gamma=-2$  &  $10.0 \pm 0.5$ & $18.2\pm 1.5$ &  0.97  &  6 -- 15.2  & 15.8         & 15.8         & 0.70 -- 1.12 \\
power-law, $\gamma=-1$  &  $25.8 \pm 3.7$ & $36.6\pm 1.2$ &  0.46  & 17.3 -- 33.4  & 33.6         & 33.6         & 0.30 -- 0.46 \\
Fiducial LOC model               &  $38.9 \pm 2.1$ & $43.5\pm 1.4$ &  0.51  & 23.9 -- 42.5  & 31.2 -- 38.4 & 35.9 -- 42.7 & 0.30 -- 0.54 \\
LOC low continuum normalisation           &  $36.3 \pm 1.9$ & $41.5\pm 1.3$ &  0.60  & 25.7 -- 40.0  & 26.5 -- 36.2 & 31.8 -- 41.1 & 0.35 -- 0.67 \\
LOC high continuum normalisation         &  $39.6 \pm 2.3$ & $44.1\pm 1.4$ &  0.43  & 23.8 -- 46.2  & 34.9 -- 39.5 & 38.0 -- 42.5 & 0.23 -- 0.45 \\
   &  &  &  &  &  &  &  \\ 
   &  &  &  &  &  &  &  \\ \hline
\multicolumn{8}{c}{$R_{\rm out} = 50$~light-days, fixed boundaries} \\ \hline
LOC model  & $22.8 \pm 0.8$  & $29.8 \pm 1.4$ & 0.42 & 21.2 -- 29.0 &  21.4 -- 23.9 & 23.1 -- 25.6  & 0.30 -- 0.49 \\
LOC model     & $17.9 \pm 1.0$  & $24.2 \pm 1.4$ & 0.53 & 18.7 -- 28.6 & 15.4 -- 18.9 & 17.7 -- 21.1 & 0.43 -- 0.67 \\ 
$\log \Phi_{\rm H} = 20$ at $R=7.5$ light-days    &  &  &  &  &  &  &  \\ 
    &  &  &  &  &  &  & \\ \hline
\multicolumn{8}{c}{Dusty BLR, $R_{\rm max} = 200$~light-days, $R_{\rm subl} = 100$~light-days, variable boundaries} \\ \hline
$\tau_{\rm subl}=\tau_{\rm cond} = 1$ days   & $57.9 \pm 3.5$ & $66.6 \pm 2.5$ & 0.81 & 38.1 -- 123.3 & 23.1 -- 49.2 & 27.1 -- 55.5 & 0.10 -- 0.84 \\
$\tau_{\rm subl}=\tau_{\rm cond}=300$~days   & $45.7 \pm 2.3$ & $53.0 \pm 1.8$ & 0.60 & 25.4 -- 52.5  & 29.7 -- 42.4 & 34.5 -- 48.0 & 0.21 -- 0.62 \\
$\tau_{\rm subl}=\tau_{\rm cond}=1000$~days  & $41.9 \pm 2.5$ & $47.1 \pm 1.4$ & 0.54 & 24.8 -- 46.3  & 31.2 -- 40.7 & 36.2 -- 46.1 & 0.26 -- 0.58 \\ \hline 
    &  &  &  &  &  &  & \\ \hline
\multicolumn{8}{c}{Dusty BLR, $R_{\rm out} = 200$~light-days, $R(\tau_{\rm subl}) = 50$~light-days, variable boundaries} \\ \hline
$\tau_{\rm subl}=\tau_{\rm cond}=300$~days   & $33.4 \pm 2.0$ & $47.3 \pm 1.7$  & 0.55 & 23.8 -- 48.1 & 22.6 -- 32.4 & 26.8 -- 38.7 & 0.24 -- 0.60 \\ \hline \hline
\multicolumn{8}{l}{$^{\star}$ CCF centroids have been calculated over
  the range in delays for which the CCF coefficient $> 0.8$ of the
  peak value. Quoted values and their }\\
\multicolumn{8}{l}{uncertainties have been
  determined using the model-independent FR/RSS Monte Carlo method
  described in Peterson et al. (1998). Each model  }\\
\multicolumn{8}{l}{light-curve is first
  sampled in the same fashion as the observations. We then compute
  1000 realisations of each light-curve, assuming random } \\
\multicolumn{8}{l}{sampling,
  with full replacement. Errors on individual data points have been drawn
  from a random Gaussian deviate with dispersion of 1\% of the}\\
\multicolumn{8}{l}{  measured flux.}\\
\multicolumn{8}{l}{$^{\dagger}$ $<\eta_{\rm eff}>$  values are here determined as in Goad et al. (2004), from the ratio $d \log F_{\rm line}/d \log F_{\rm cont}$.} \\
\multicolumn{8}{l}{$^{\ddagger}$ Measured ranges have been determined from measurements of each of the 13 seasons of data for NGC~5548.}\\
\end{tabular}
\end{table}

\twocolumn

\section{Acknowledgements}
\thanks{

We would like to thank the anonymous referee for providing many
insightful suggestions which have led to improved clarity of the work
presented here. Mike Goad would like to thank the generous hospitality
of Kirk and Angela Korista and the Department of Physics \& Astronomy
at Western Michigan University during the initial stages of this
work. Kirk Korista would like to thank the University of Leicester for
their hospitality during the completion of this work.}

\section{References}


\noindent Ade, et al. 2014, A\&A 61, A16.

\noindent Baldwin, J. Ferland, G., Korista, K., and Verner, D. 1995, ApJ 455, 119.

\noindent Barth, A.J., Pancoast, A., Bennert, V.N. et al. 2013, ApJ 769, 128. 

\noindent Barth, A.J., Pancoast, A., Thorman, S.J.  et al. 2011, ApJ 743, 4. 

\noindent Barvainis, R. 1987, ApJ 320, 537.

\noindent Bentz, M.C., Denney, K.D., Grier, C.J. et al. 2013, ApJ 767, 149.

\noindent Bentz, M.C., Horne, K., Barth, A.J. et al. 2010 ApJ 720, 46.


\noindent Bentz, M. and Katz, S. 2015, PASP 127, 67.

\noindent Bentz, M., Peterson, B.M., Netzer, H., Pogge, R.W, and Vestergaard, M. 2009a, ApJ 697, 160.

\noindent Bentz, M.C., Walsh, J.L., Barth, A,J. et al. 2009b, ApJ 705, 199

\noindent Bentz, M.,  Walsh, J.L., Barth, A,J. et al. 2008, ApJ 689, 21.

\noindent Bentz, M.C., Denney, K.D., Cackett, E.M. et al. 2007, ApJ 662, 205.

\noindent Bentz, M.C., Peterson, B.M., Pogge, R.W. et al. 2006, ApJ 644, 133.

\noindent Blandford, R.D. and McKee, C.F. 1982, ApJ 255, 419.

\noindent Bottorff, M.C., Baldwin, J.A., Ferland G.J. et al. 2002, ApJ 581, 932.

\noindent Bottorff, M., Ferland G., Baldwin, J. and Korista, K. et al. 2000, ApJ 542, 644.

\noindent Cackett, E.M. and Horne, K. 2006, MNRAS 365, 1180.

\noindent Cackett, E.M., Gultekin, K., Bentz, M. C., et al. 2015, arXiv:1503.02029.

\noindent Chiang, J., Reynolds, C.S., Blaes, O.M. et al. 2000, ApJ 528, 292.

\noindent Clavel, J. Reichert, G.A. Alloin, D. et al. 1991, ApJ 366, 64.



\noindent Collin.-Souffrin, S. 1986 A\&A 166, 115.

\noindent Collin, S., Kawaguchi, T., Peterson, B.M. and Vestergaard, M. 2006 A\&a 456, 75. 


\noindent Davidson, K. and Netzer, H. 1979, Rev. Mod. Phys. 51, 715.

\noindent Denney, K. D., Peterson, B. M., Pogge, R. W, et al. 2010, ApJ 721, 715.

\noindent Denney, K. D., Peterson, B. M., Pogge, R. W, et al. 2009, ApJ 704, 80.

\noindent Kilerci Eser, E. Vestergaard, M. Peterson, B.M. et al. 2015, ApJ 801, 8.


\noindent Ferland, G.J., Korista, K. T., Verner, D. A. et al. 1998, PASP 110, 761

\noindent Ferland, G. J., Korista, K. T. and  Verner, D. A. 1997, In:
Astronomical Data Analysis Software and Systems VI, A.S.P. Conference Series, Vol. 125, 1997, Gareth Hunt and H. E. Payne, eds., p. 213.


\noindent Ferland, G. J., Korista, K. T., Verner, D. A., Ferguson, J. W., Kingdon, J. B. and Verner, E. M. 1998, In:  The Publications of the Astronomical Society of the Pacific, Volume 110, Issue 749, pp. 761-778.

\noindent Ferland, G.J., Peterson, B.M., Horne, K. Welsh, W.F., Nahar, S.N. 1992, ApJ 387, 95.

\noindent Ferland, G. and Netzer, H. 1979, ApJ 229, 274.

\noindent Gaskell, C.M. and Peterson, B.M. 1987, ApJS 65, 1.

\noindent Gilbert, K.M. and Peterson, B.M. 2003, ApJ 587, 123.

\noindent Goad, M.R. and Korista, K.T. 2014, MNRAS 444, 43.

\noindent Goad, M. R.; Korista, K. T. and Ruff, A.J. 2012, MNRAS 426, 3086.

\noindent Goad, M. R.; Korista, K. T.; Knigge, C., 2004, MNRAS 352, 277.



\noindent Grier, C., Peterson, B.M., Horne, K. et al. 2013, ApJ 764, 47.

\noindent H\"{o}nig, S.F. and Kishimoto, M. 2011, A\&A 534, 121.

\noindent Horne, K., Korista, K.T. and Goad, M.R. 2003, MNRAS 339, 367.

\noindent Horne, K., Welsh, W.F. and Peterson, B.M. 1991, ApJ 367, 5.

\noindent Kong, M.-Z., Wu, X.-B, Wang, R. Liu, F.K. and Han, J.L. 2006, A\&A 456, 473.

\noindent Kaspi, S. and Netzer, H. 1999, ApJ 524, 71.


\noindent Kelly, B.C., Bechtold, J., Siemiginowska, A. 2009, ApJ 698, 895.

\noindent Kinney, A.L., Rivolo, A.R. and Koratkar, A.P. 1990, ApJ 357, 338.

\noindent Kishimoto, M., H\"{o}nig, S.F., Antonucci, R. et al. 2013 ApJ 775, 36.

\noindent Kishimoto, M., H\"{o}nig, S. F., Antonucci, R., Millour, F. Tristram, K. R. W., and Weigelt, G. 2011, A\&A 536, 78.

\noindent Korista, K., Ferland, G. and Baldwin, J. 1997, ApJ 487, 555.

\noindent Korista, K., Baldwin, J., Ferland, G., and Verner, D. 1997, ApJS 108, 401.

\noindent Korista, K.T., Alloin, D., Barr, P. et al. 1995, ApJS 97, 258.

\noindent Korista, K.T. and Goad, M.R. 2004, ApJ 606, 749.

\noindent Korista, K.T. and Goad, M.R. 2001, ApJ 553, 695.

\noindent Korista, K.T. and Goad, M.R. 2000, ApJ 536, 284.

\noindent Koshida, S., Minezaki, T., Yoshii, Y. et al. 2014, ApJ 788, 159.

\noindent Koshida, S., Yoshii, Y., Kobayashi. Y. et al. 2009, ApJ 700, 109.


\noindent Krause, M., Schartmann, M. and Burkert, A. 2012, MNRAS 425, 3172.

\noindent Krolik, J.H., Horne, K., Kallman, T.R. et~al. 1991, ApJ 371, 541.

\noindent Kuehn, C.A., Baldwin, J.A., Peterson, B.M. and Korista, K.T. 2008, ApJ 673, 69.

\noindent MacLeod, C.L., Ivezi\'{c}, Z., Kochanek,C.S. et al. 2010, ApJ 721, 1014.


\noindent Maoz, D., Netzer, H., Peterson, B.M. et al. 1993, ApJ 404, 576.

\noindent Marshall, H.L., Carone, T.E., Peterson, B.M. et al. 1997, ApJ 479, 222.

\noindent Mathews, W.G. and Ferland, G.J. 1987, ApJ 323, 456.

\noindent Mathews, W.G. 1982, ApJ 252, 39.

\noindent McLure, R.J. and Jarvis, M.J. 2002, MNRAS 337, 109.


\noindent Mor, R., and Trakhtenbrot, B. 2011, ApJ 737, L36.

\noindent Mor, R. and Netzer, H. 2012, MNRAS 420, 526.

\noindent Mushotzky, R.F., Edelson, R., Baumgartner, W. and Gandhi, P, ApJ 2011, 743, 12.

\noindent Nemmen, R.S. and Brotherton, M.S. 2010, MNRAS 408, 1598.

\noindent Nenkova, M., Sirocky, M.M., Nikutta, R. et al. 2008, ApJ 685, 160.

\noindent Netzer, H. and Laor, A. 1993, ApJ 404, 51.

\noindent Netzer, H. 1985, ApJ 289, 451.


\noindent Netzer, H. 1987, MNRAS, 225, 55.

\noindent O'Brien, P.T. Goad, M.R. and Gondhalekar,  P.M. 1994, MNRAS 268, 845.



\noindent Pancoast, A., Brewer, B.J., Treu, T. et al. 2012, ApJ 754, 49..

\noindent Pancoast, A., Brewer, B.J., Treu, T. et al. 2014a MNRAS 445, 3073

\noindent Pancoast, A., Brewer, B.J., Treu, T. 2014b MNRAS, 445, 3055.


\noindent P\'{e}rez, E, Robinson, A. and de La Fuente, L. 1992, MNRAS 256, 103.


\noindent Peterson, B.M., Denney, K.D., De Rosa, G. et al. 2013, ApJ 2013, 779, 109.

\noindent Peterson, B.M., Ferrarese, L., Gilbert, K. M., Kaspi, S., Malkan, M. A., et al. 2004, ApJ 613, 682.

\noindent Peterson, B.M. Berlind, P. Bertram, R. Bischoff, K., Bochkarev, N.G. et al. 2002 ApJ 581, 197.

\noindent Peterson, B.M. and Wandel, A. 2000, ApJ 540, 13.

\noindent Peterson, B.M., Wanders, I., Horne, K, et al. 1998, PASP 110, 660. 

\noindent Pogge, R.W. and Peterson, B.M. 1992, AJ, 103, 1084.




\noindent Romanishin, W., Balonek, T.J. Ciardullo, R. et al. 1995, ApJ 455, 516.

\noindent Schlafly, E.F. and Finkbeiner, D.P. 2011, ApJ 737, 103.

\noindent Schn\"{u}lle, K., Pott, J.-U., Rix, H.-W. et al. 2013, A\&A 557, L13.

\noindent Schn\"{u}lle, K., Pott, J.-U., Rix, H.-W. et al. 2015,  A\&A 578, 57.

\noindent Skielboe, A., Pancoast, A., Treu, T. et al. 2015, arXiv:1502.02031.


\noindent Suganuma, M., Yoshii, Y., Kobayashi, Y. et al. 2006, ApJ 639, 46.

\noindent Suganuma, M., Yoshii, Y., Koboyashi, Y. et al. 2004, ApJ 612, 113.





\noindent Vestergaard, M and Peterson, B.M. 2006, ApJ 641, 689.

\noindent Vestergaard, M and Peterson, B.M. 2005, ApJ 625, 688.

\noindent Welsh, W.F. and Horne, K. 1991, ApJ 379, 586.



\appendix
\label{lastpage}

\end{document}